\documentclass[%
 reprint,
unsortedaddress,
 amsmath,amssymb,
 aip,
]{revtex4-2}

\usepackage{hyperref}
\usepackage{physics}
\hypersetup{colorlinks=true, allcolors=blue}
\usepackage{comment}

\usepackage{graphicx}
\usepackage{dcolumn}
\usepackage{bm}
\usepackage{textcomp}
\usepackage{amsmath}
\def\mathclap#1{\text{\hbox to 0pt{\hss$\mathsurround=0pt#1$\hss}}}

\newcommand*{\citen}[1]{%
  \begingroup
    \romannumeral-`\x 
    \setcitestyle{numbers}%
    \citep{#1}%
  \endgroup   
}

\makeatletter
\newcommand{\vast}{\bBigg@{5}}
\newcommand{\Vast}{\bBigg@{5}}
\makeatother

\usepackage{xcolor}

\newcommand*{\colorboxed}{}
\def\colorboxed#1#{%
  \colorboxedAux{#1}%
}

\usepackage{tikz}
\usepackage{booktabs}
\usetikzlibrary{calc}

\usepackage{newfloat}
\DeclareFloatingEnvironment[
    fileext=los,
    listname={List of Schemes},
    name=MOV.,
    placement=tbhp,
]{movie}

\newcommand*{\colorboxedAux}[3]{%
  \begingroup
    \colorlet{cb@saved}{.}%
    \color#1{#2}%
    \boxed{%
      \color{cb@saved}%
      #3%
    }%
  \endgroup
}

\usepackage{subfigure}

\usepackage{multirow}
\usepackage{xcolor}
\usepackage{soul}
\begin{document}

\preprint{APS/123-QED}

\title[Mn$_3$Sn with strain]{Impact of strain on the SOT-driven dynamics of thin film Mn$_3$Sn}
\author{Ankit Shukla}
 \email{ankits4@illinois.edu}
 \author{Siyuan Qian}
 \email{siyuanq3@illinois.edu}
\author{Shaloo Rakheja}
 \email{rakheja@illinois.edu}
\affiliation{Holonyak Micro and Nanotechnology Laboratory, University of Illinois at Urbana-Champaign, Urbana, IL 61801}

\date{\today}

\begin{abstract}
Mn$_3$Sn, a metallic antiferromagnet with an anti-chiral 120$^\circ$ spin structure, generates intriguing magneto-transport signatures such as a large anomalous Hall effect, spin-polarized current with novel symmetries, anomalous Nernst effect, and magneto-optic Kerr effect. When grown epitaxially as MgO(110)[001]$\parallel$ Mn$_3$Sn($0\bar{1}\bar{1}0$)[0001], Mn$_3$Sn experiences a uniaxial tensile strain, which changes the bulk six-fold anisotropy to a two-fold perpendicular magnetic anisotropy (PMA). Here, we investigate the field-assisted spin orbit-torque (SOT)-driven dynamics in single-domain Mn$_3$Sn with PMA. We find that for non-zero external magnetic fields, the magnetic octupole \textcolor{black}{moment} of Mn$_3$Sn can be switched between the two stable states if the input current is between two field-dependent critical currents. Below the lower critical current, the magnetic octupole \textcolor{black}{moment} exhibits a stationary state in the vicinity of the initial stable state. On the other hand, above the higher critical current, the magnetic octupole \textcolor{black}{moment} shows oscillatory dynamics which could, {in principle}, be tuned from the 100s of megahertz to the {terahertz} range. We obtain approximate {analytic} expressions of the two critical currents that agree very well with the numerical simulations for experimentally relevant magnetic fields. We also obtain unified functional form of the switching time versus the input current for different magnetic fields. Finally, we show that for lower values of Gilbert damping ({$\alpha \lesssim 2\times 10^{-3}$}), the critical currents and the final steady-states depend significantly on $\alpha$. The numerical and analytic results presented in our work can be used by both theorists and experimentalists to understand the SOT-driven order dynamics in PMA Mn$_3$Sn and design future experiments and devices.
\end{abstract}

\maketitle

\section{Introduction}
\vspace{-5pt}
Antiferromagnets (AFMs) are a class of magnetic materials that produce negligible stray fields, are robust to external magnetic field perturbations, and exhibit resonant frequency in the terahertz (THz) regime. 
These distinctive properties are a consequence of strong exchange interactions between the uniquely arranged spins of the neighboring atoms, and a negligible net macroscopic magnetization.~\citep{gomonay2014spintronics, jungwirth2016antiferromagnetic, baltz2018antiferromagnetic, jungfleisch2018perspectives}  
AFMs are, therefore, considered as promising candidates for building next generation magnonic devices, high-density memory devices, and ultrafast signal generators.~\cite{han2023coherent}
Among the various possible AFMs, noncollinear but coplanar metallic AFMs of the form Mn$_3$X, with a triangular spin structure, have recently been explored extensively owing to their intriguing magneto-transport characteristics, such as a large spin Hall effect (SHE),~\citep{zhang2016giant} anomalous Nernst effect (ANE), anomalous Hall effect (AHE)~\citep{kubler2014non, zhang2017strong, iwaki2020large, tsai2021large} and magneto-optical Kerr effect (MOKE),~\citep{higo2018large} ferromagnet-like spin-polarized currents,~\citep{vzelezny2017spin, wang2023spin} and a finite tunneling magnetoresistance (TMR).~\citep{qin2023room, chen2023octupole}
These noncollinear AFMs are chiral in nature and could be further classified as positive (X = Ir, Pt, Rh) or negative (X = Sn, Ge, Ga) chirality materials based on the type of spin interaction.~\citep{yamane2019dynamics}

Here, we focus on thin-film Mn$_3$Sn, owing to its various technologically-relevant properties.
Bulk Mn$_3$Sn, which is a six-fold spin-degenerate chiral antiferromagnet with a small net magnetization, has a high N\'eel temperature of approximately $420-430~\mathrm{K}$.~\citep{tomiyoshi1982magnetic, sung2018magnetic, higo2018large}
Recent experiments have demonstrated that the magnetic order parameter in Mn$_3$Sn, referred to as the magnetic octupole moment, can be switched between the six stable states using spin-orbit torque (SOT) in a bilayer setup of AFM and heavy metal (HM).~\citep{tsai2020electrical, takeuchi2021chiral, pal2022setting, krishnaswamy2022time, xu2023robust}
The critical charge current density required to switch the magnetic octupole \textcolor{black}{moment} was found to be of the order of $10^{6}-10^{7}~\mathrm{A/cm^2}$, which is smaller than or comparable to that required to switch the magnetization in most ferromagnets ($\sim10^7-10^{8}~\mathrm{A/cm^2}$).~\citep{ramaswamy2018recent}
Some experiments have also alluded to current-driven oscillations of the octupole moment, with frequencies in the range of 100's of MHz to a few GHz, in the bilayer setup.~\citep{takeuchi2021chiral, yan2022quantum}
In all of these experiments, the changes in the octupole moment were detected via the AHE since Mn$_3$Sn exhibits large anomalous Hall conductivity, ranging between $30-40~\mathrm{\Omega^{-1}~cm^{-1}}$ at $300~\mathrm{K}$, owing to the broken time-reversal symmetry (TRS).~\citep{kubler2014non, markou2018noncollinear, liu2023anomalous}
The magnitude and sign of the AHE signal can be further modulated by the application of small in-plane tensile or compressive uniaxial strain of the order of $0.1\%$, as revealed recently.~\citep{ikhlas2022piezomagnetic}
Such uniaxial strains alter the crystal symmetry, followed by the spin degeneracy, and hence the Hall conductivity.~\citep{ikhlas2022piezomagnetic, dasgupta2022tuning}

Thin films of Mn$_3$Sn, when grown epitaxially on MgO(110)[001] substrate, also experience in-plane tensile strain, arising from the lattice mismatch between Mn$_3$Sn and MgO.
Consequently, the six-fold spin-degenerate system reduces to a two-fold spin-degenerate system, with a comparatively larger net magnetization, leading to perpendicular magnetic anisotropy (PMA) in such films.~\citep{higo2022perpendicular, yoon2023handedness}
\textcolor{black}{The AHE measurements further revealed that the magnetic octupole moment of the PMA Mn$_3$Sn films, used in a bilayer setup, can be deterministically switched in the presence of a symmetry-breaking magnetic field, which is parallel to the current direction.~\citep{higo2022perpendicular, yoon2023handedness}
Recently, all antiferromagnetic tunnel junctions comprising Mn$_3$Sn/MgO/Mn$_3$Sn}, utilizing PMA Mn$_3$Sn, were found to display non-zero TMR of about $2\%$ at room temperature\textcolor{black}{---}owing to the TRS breaking and the momentum-dependent spin splitting of electronic band structure.~\citep{chen2023octupole, dong2022tunneling}
These promising developments make thin-film PMA Mn$_3$Sn a strong candidate for future high-density memory and ultrafast nano-oscillator devices.

For a better understanding of the current-driven dynamics, on the theoretical front, Higo~\emph{et al.} presented a brief numerical investigation of the different possible steady-states in PMA Mn$_3$Sn, as a function of the applied current and magnetic field.~\citep{higo2022perpendicular}
Their study, however, is limited in its scope as it does not provide an insight into the dependence of the different dynamical regimes on the intrinsic energy scale of Mn$_3$Sn, or its material parameters.
Analytic expressions of threshold currents, switching times, and oscillation frequency as functions of material parameters and input stimuli are also lacking. 
Another recent work, focusing on Mn$_3$Sn with uniaxial strain, numerically investigated only the field-free oscillation and pulsed-SOT switching dynamics.~\citep{dasgupta2022tuning}
In their very recent work, Yoon \textcolor{black}{and Zhang}~\emph{et al.} developed analytic models of the first- and second-harmonic Hall resistances and successfully validated them against experimental observation.~\citep{yoon2023handedness} Their analysis, however, was limited to the quasi-static regime.
Previous theoretical works have also explored current-driven switching and oscillation dynamics in AFMs with two-fold spin degeneracy.~\citep{gomonay2012symmetry, gomonay2015using, shukla2022spin} However, those materials were not representative of Mn$_3$Sn with uniaxial strain since the net magnetization was considered zero. 
In this work, therefore, we address the existing limitations and investigate, both numerically and analytically, the magnetic field-assisted SOT-driven deterministic switching and oscillation dynamics in monodomain Mn$_3$Sn, with two stable states.

For the numerical investigation of the static and dynamic properties of single-domain Mn$_3$Sn with in-plane uniaxial strain, an energy interaction model is presented in Section~\ref{sec:energy} of this work.
To elucidate the properties of the ground states as well as their dependence on the material parameters, a simpler analytic model of the energy interaction is perturbatively obtained in Section~\ref{sec:energy_pertur}, and shown to agree well with the numerical results. 
Next, the field-assisted SOT-driven dynamics of the magnetic order in PMA Mn$_3$Sn is analyzed in Section~\ref{sec:SOT}.
Analytic models, pertinent to the current-driven dynamics such as threshold currents, stationary states, switching time and oscillation frequency, are presented in detail.
Previously, we had utilized this framework to analyze field-free SOT-driven dynamics in monodomain Mn$_3$Sn with six states~\citep{shukla2023order}.
In our present work, however, we build models that elucidate the impact of strain as well as magnetic field on the dynamics.
The impact of the Gilbert damping constant on the dynamics is investigated and the salient features are discussed in Section~\ref{sec:damping}.
The field-assisted SOT-driven dynamics in Mn$_3$Sn with \textcolor{black}{no strain}~\cite{xu2023deterministic} and compressive strain is discussed in supplementary material, along with a brief discussion of the AHE and the TMR detection schemes.

\section{Free Energy Model and Ground States}\label{sec:energy}
\begin{figure}[ht!]
  \centering
  \includegraphics[width = \columnwidth, clip = true, trim = 0mm 0mm 0mm 0mm]{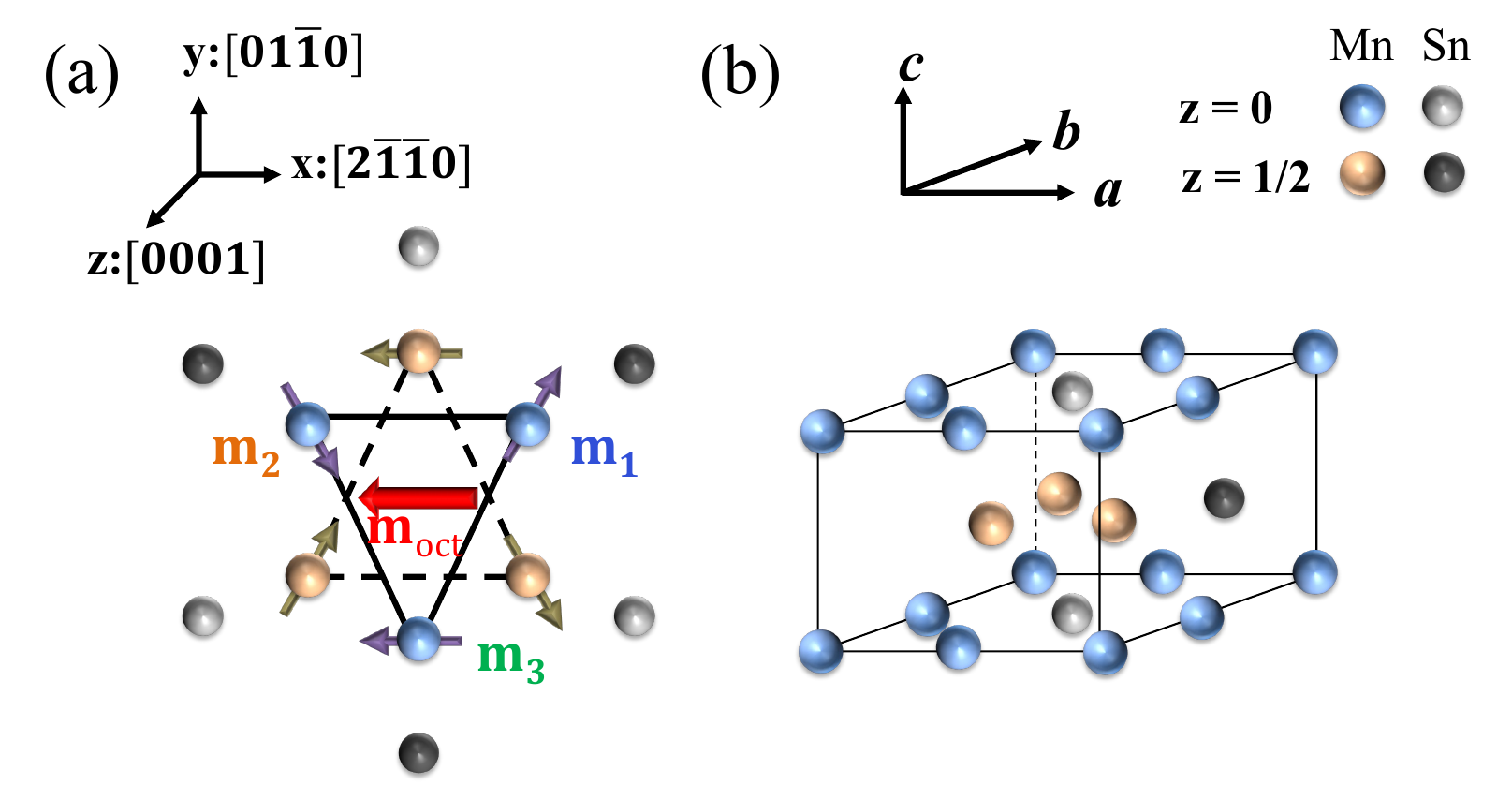}
  \caption{(a) Atomic and one of the six possible spin arrangements in the basal plane of bulk Mn$_3$Sn. (b) Unit cell of Mn$_3$Sn.} 
  \label{fig:crystal}
\end{figure}

\vspace{-5pt}
Below its N\'eel temperature, Mn$_3$Sn crystallizes into a hexagonal Kagome $D0_{19}$ lattice and can only be stabilized in slight excess of Mn atoms. 
The Mn atoms are located at the corners of the hexagons whereas the Sn atoms are located at their respective centers.
Such lattices are stacked along the $c$ axis $([0001]$ direction$)$ in an ABAB arrangement. A simple representation of the crystal structure is presented in Fig.~\ref{fig:crystal}. 
In {each} Kagome plane, the magnetic moments on the Mn atoms form a geometrically frustrated noncollinear triangular spin structure (Fig.~\ref{fig:crystal}(a)), with spins on the nearest neighbors aligned at an angle of approximately $120^\circ$ with respect to each other.~\citep{tomiyoshi1982magnetic}
These spins are canted slightly toward the in-plane easy axes, resulting in a small net magnetization, which is six-fold degenerate in the Kagome plane.~\citep{markou2018noncollinear}
Under the application of a small in-plane uniaxial strain, the system becomes two-fold degenerate, and its net magnetization is also altered.~\citep{ikhlas2022piezomagnetic, higo2022perpendicular, yoon2023handedness}


To investigate the static and the current-driven dynamic behavior of a single-domain particle of uniaxially strained-Mn$_3$Sn, {comprising} three interpenetrating sublattices, the free energy density is defined as~\citep{ikhlas2022piezomagnetic, higo2022perpendicular}
\begin{flalign}\label{eq:energy_density}
    \begin{split}
        &F\qty(\vb{m}_1, \vb{m}_2, \vb{m}_3) =  J_E\left(\qty(1+\delta_E)\vb{m}_{1} \vdot \vb{m}_{2} + \vb{m}_{2} \vdot \vb{m}_{3} \right. \\
        &\left.+ \vb{m}_{3} \vdot \vb{m}_{1} \right) + D_M \vb{z} \vdot \left(\vb{m}_{1} \cp \vb{m}_{2} + \vb{m}_{2} \cp \vb{m}_{3} \right. \\
        &\left.+ \vb{m}_{3} \cp \vb{m}_{1} \right) - \sum_{i = 1}^{3}\qty(K_{e} \qty(\vb{m}_{i} \vdot \vb{u}_{e, i})^{2} + M_{s}\vb{H}_a \vdot \vb{m}_{i}),
    \end{split}
\end{flalign}
where $\vb{m}_{1}, \vb{m}_{2}$, and $\vb{m}_{3}$ are the magnetization vectors corresponding to the three sublattices, while $J_E (>0)$, $D_M (>0)$, and $K_{e} (>0)$ are the symmetric exchange interaction constant, asymmetric Dzyaloshinskii-Moriya interaction (DMI) constant, and single-ion uniaxial magnetocrystalline anisotropy constant, respectively.
Each magnetization vector has a constant saturation magnetization, $M_s$.
Here, it is assumed that the uniaxial strain acts \textcolor{black}{between $\vb{m}_1$ and $\vb{m}_2$}. The effect of this uniaxial strain is included in the empirical parameter $\delta_E$---a positive (negative) value indicates a stronger (weaker) exchange interaction between $\vb{m}_1$ and $\vb{m}_2$.~\citep{ikhlas2022piezomagnetic, higo2022perpendicular, yoon2023handedness}
Therefore, a positive (negative) $\delta_E$ corresponds to a shorter (longer) bond length and hence compressive (tensile) strain.
The last term in Eq.~(\ref{eq:energy_density}) \textcolor{black}{represents} the Zeeman energy due to the externally applied magnetic field $\vb{H}_a$. 
Finally, $\vb{u}_{e, i}$ is the local easy axis corresponding to $\vb{m}_i$. The easy axes are \textcolor{black}{assumed to be} $\vb{u}_{e, 1} = -(1/2) \vb{x} + (\sqrt{3}/2) \vb{y}$, $\vb{u}_{e, 2} = -(1/2) \vb{x} - (\sqrt{3}/2) \vb{y}$ and $\vb{u}_{e, 3} = \vb{x}$.
Mn$_3$Sn is an exchange dominant AFM such that $J_E \gg D_M \gg K_e$.
Typical values of the material parameters of Mn$_3$Sn, considered in this work, are listed in Table~\ref{tab:mat_params1}.
\textcolor{black}{The ground states for different $\delta_E$ and $\vb{H}_a$ can be obtained by minimizing Eq.~(\ref{eq:energy_density}) with respect to $\vb{m}_{i}$.}

\begin{figure}[ht!]
  \centering
  \includegraphics[width = 0.85\columnwidth, clip = true, trim = 20mm 15mm 0mm 0mm]{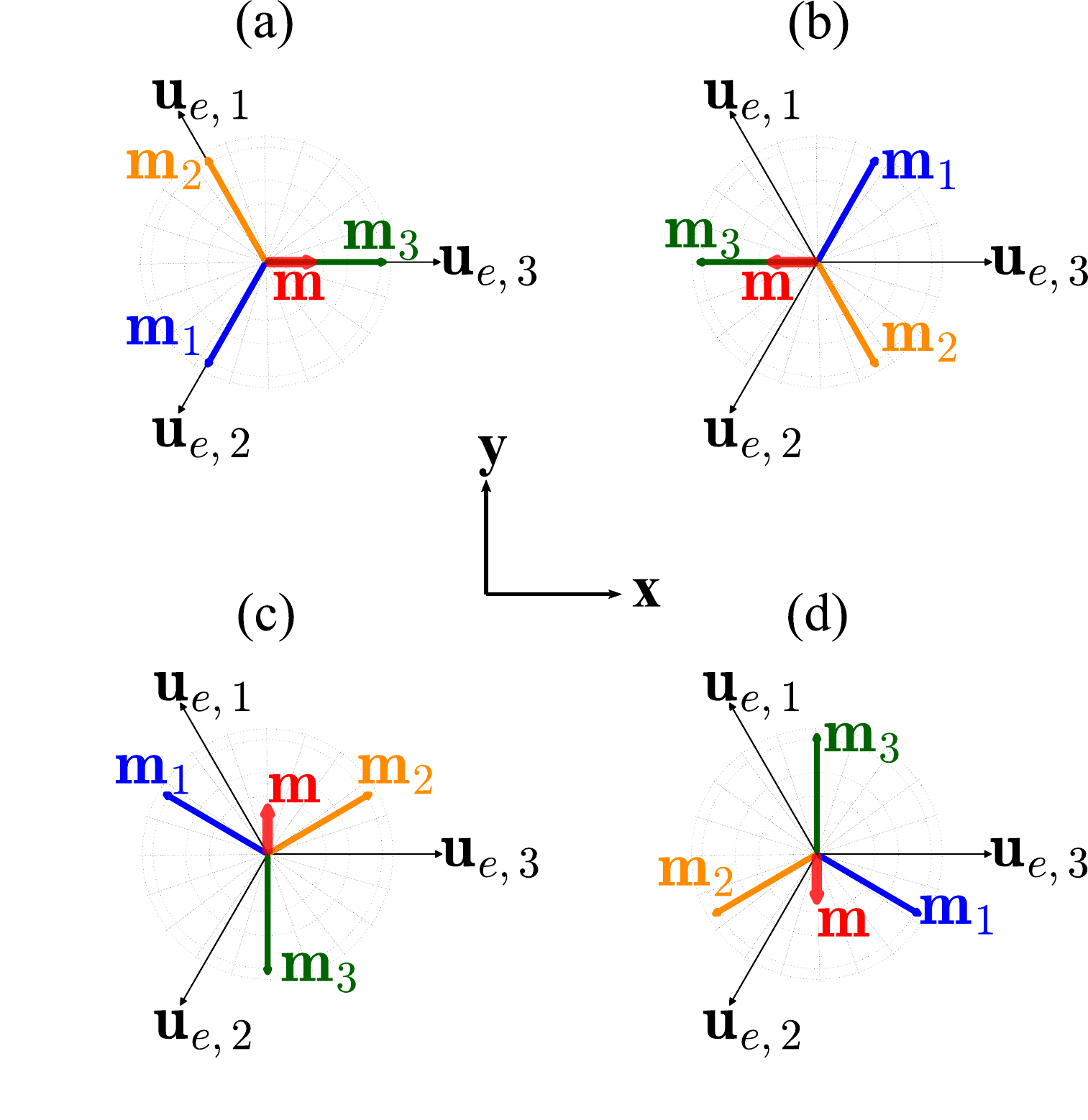}
  \caption{Equilibrium states in single-domain Mn$_3$Sn crystal under (a, b) compressive and (c, d) tensile strains.
  The sublattice vectors lie in the Kagome plane which is assumed to coincide with the x-y plane.
  (a, b) Only $\vb{m}_3$ coincides with its easy axis. 
  (c, d) None of the sublattice vectors coincide with their local easy axes. Instead, tensile strain forces $\vb{m}_3$ perpendicular to its easy axis.
  The two equilibrium states for compressive as well as tensile strains are separated by $180^\circ$ with respect to each other.
  A small \textcolor{black}{in-(Kagome)-plane} average magnetization, $\vb{m}$, exists parallel (antiparallel) to $\vb{m}_3$ in the case of compressive (tensile) strain.
  The magnitude of $\vb{m}$ depends on the strength of the strain.
  Here, $\vb{m}$ is not drawn to scale but magnified by $100 \times$ for the purpose of clear representation.} 
  \label{fig:equilibrium}
\end{figure}

\textcolor{black}{The ground states of single-domain strained-Mn$_3$Sn, in the absence of any magnetic field $\qty(\vb{H}_a = 0)$, are shown in Fig.~\ref{fig:equilibrium}.
In all the cases, $\vb{m}_1$, $\vb{m}_2$, and $\vb{m}_3$ exhibit a clockwise ordering with approximately 120$^\circ$ angle between them, all lying within the x-y plane.
Compared to the six-fold degeneracy observed in single-domain Mn$_3$Sn with no strain,~\citep{liu2017anomalous, shukla2023order} a two-fold spin degeneracy is observed in strained-Mn$_3$Sn, where $\vb{m}_3$ coincides with its easy axis in the case of compressive strain (Fig.~\ref{fig:equilibrium}(a, b)) while it is perpendicular to its easy axis in the case of tensile strain (Fig.~\ref{fig:equilibrium}(c, d)).}
A small non-zero net magnetization, $\vb{m} = \frac{\vb{m}_1 + \vb{m}_2 + \vb{m}_3}{3}$, exists in single-domain Mn$_3$Sn with no strain.~\citep{liu2017anomalous, shukla2023order}
We find that strained-Mn$_3$Sn also hosts a net magnetization which is parallel (antiparallel) to $\vb{m}_3$ in the case of the compressive (tensile) strain. 
For $|\delta_E| = 10^{-3}$, which represents a uniaxial strain of $0.1\%$, the norm of the net magnetization increases from the bulk value of $\norm{\vb{m}} \approx 3.66 \times 10^{-3}$ to $\norm{\vb{m}} \approx 3.95 \times 10^{-3}$ $\qty(\norm{\vb{m}} \approx 3.92 \times 10^{-3})$ in the case of compressive (tensile) strain. 
\textcolor{black}{The non-zero $\vb{m}$ suggests that the angle between the sublattice vectors is not exactly $120^\circ$.
Indeed,} in the case of compressive (tensile) strain, $\eta_{12} \approx 0.78^\circ \qty(-0.78^\circ)$ and $\eta_{23} = \eta_{31} \approx -0.39^\circ \qty(0.39^\circ)$, where
 $\eta_{ij} = \cos^{-1}{(\vb{m}_i \vdot \vb{m}_j)} - \frac{2\pi}{3}$.
For both compressive and tensile strain, $\eta_{12} + \eta_{23} + \eta_{31} = 0$.
\textcolor{black}{For $\delta_E = 0$, our calculations showed that the respective $|\eta_{ij}|$ were different from the values reported above and depended on the ground state under consideration.~\cite{shukla2023order}
The aforementioned results with and without strain are as expected---strong exchange and DM interactions attempt enforcing a clockwise ordering of $\vb{m}_1$, $\vb{m}_2$, and $\vb{m}_3$ with exactly 120$^\circ$ angle between them, within a plane perpendicular to the z-axis. However, the anticlockwise ordering of $\vb{u}_{e, 1}$, $\vb{u}_{e, 2}$, and $\vb{u}_{e, 3}$ forces the sublattice vectors in the x-y plane with a small deviation from the 120$^\circ$ ordering. Uniaxial strain leads to further increase in the deviation.}

\begin{table}[ht!]
\caption{\label{tab:mat_params1} {List of material parameters for the AFM, Mn$_3$Sn, and the heavy metal (HM), which is chosen as W, in the SOT device setup.}}
\begin{ruledtabular}
\begin{tabular}{cccc}
Parameters  & Definition  &  Values & Ref.\\
\hline
$J_E~(\mathrm{J/m^3})$ & Exchange constant &$2.4 \times 10^8$ & \citen{yamane2019dynamics}\\
$D~(\mathrm{J/m^3})$  & DMI constant &$2 \times 10^7 $ & \citen{yamane2019dynamics}\\
$K_e~(\mathrm{J/m^3})$ & Uniaxial anisotropy constant &$3 \times 10^6$  & \citen{yamane2019dynamics} \\
$M_s~(\mathrm{A/m})$ & Saturation magnetization &$1.3 \times 10^6$ &\citen{yamane2019dynamics}\\
$|\delta_E|$ & Strain parameter & $10^{-3}$ &\citen{higo2022perpendicular} \\
$\alpha$ & Gilbert damping & $0.003$ & \citen{tsai2020electrical}\\
$\theta_\mathrm{SH}$ & Spin Hall angle for HM & $0.06$ & \citen{higo2022perpendicular}\\
\end{tabular}
\end{ruledtabular}
\end{table}

\begin{figure}[ht!]
  \centering
  \includegraphics[width = 0.85\columnwidth, clip = true, trim = 20mm 0mm 0mm 0mm]{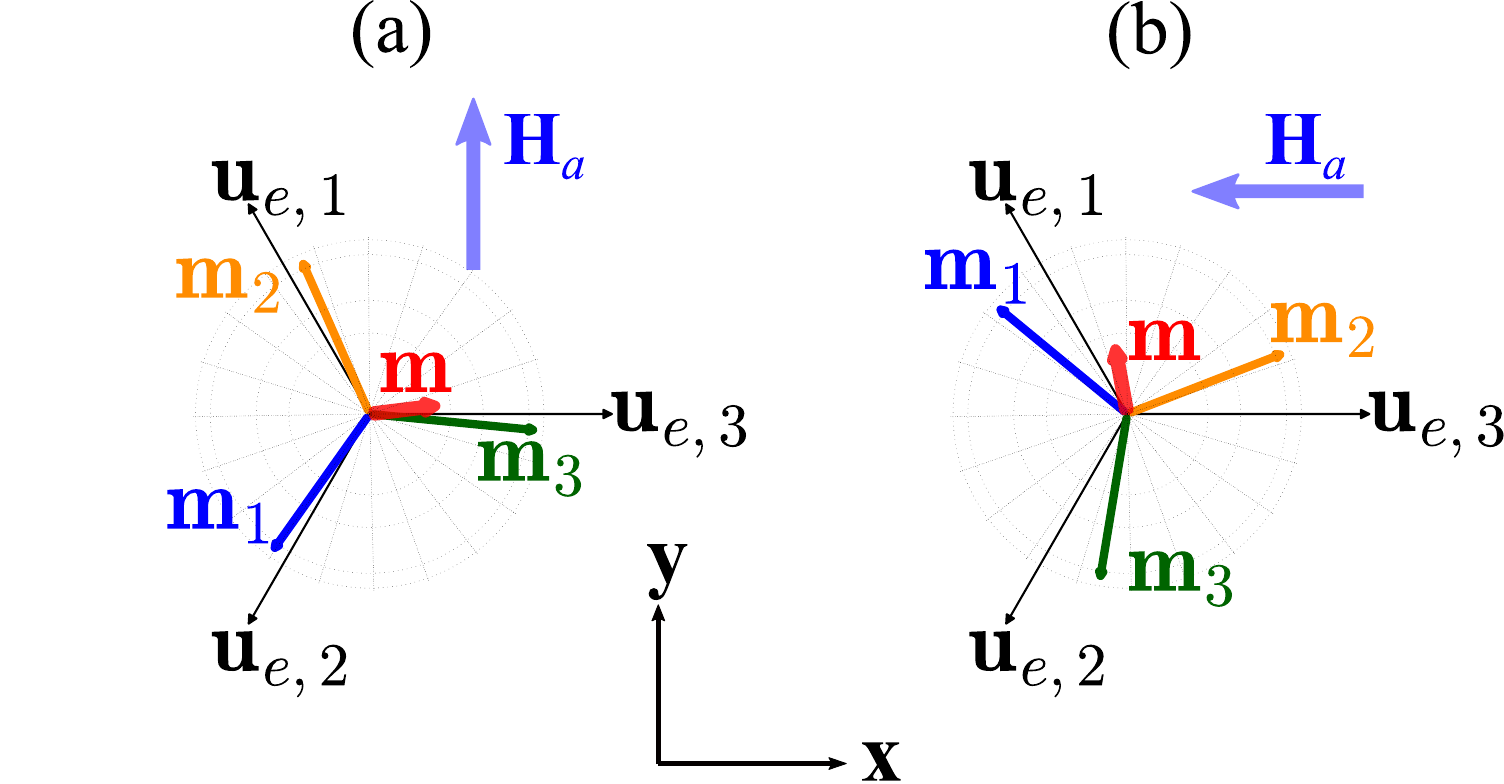}
  \caption{Ground states of monodomain Mn$_3$Sn crystal, with (a) compressive strain and (b) tensile strain, under the effect of in-plane external magnetic field, $\vb{H}_a = H_0 \qty(\cos{(\varphi_H)},\sin{(\varphi_H)}, 0)$.
  In both the cases $\vb{m}$ tilts towards $\vb{H}_a$. 
  Here, $\norm{\vb{H}_a} = 0.1~\mathrm{T}$ while $\vb{m}$ is magnified by $100 \times$ for the purpose of clear representation in both the cases.} 
  \label{fig:gs}
\end{figure}

External magnetic fields, when applied to a thin film of Mn$_3$Sn, change the energy of the system, and therefore, the ground states. Here, we only consider an external magnetic field in the Kagome plane as $\vb{H}_a = H_0 \qty(\cos{(\varphi_H)},\sin{(\varphi_H)}, 0)$, where $\varphi_{H}$ is the angle between the magnetic field and the x-axis.
Figure~\ref{fig:gs}(a) shows the ground states of Mn$_3$Sn when $\vb{H}_a = \qty(0, 0.1~\mathrm{T}, 0)$ is applied to the equilibrium state of Fig.~\ref{fig:equilibrium}(a) whereas Fig.~\ref{fig:gs}(b) shows the ground state when $\vb{H}_a = \qty(-0.1~\mathrm{T}, 0, 0)$ is applied to the equilibrium state of Fig.~\ref{fig:equilibrium}(c).
In both the cases $\vb{m}$ tilts towards the magnetic field, while the sublattice vectors either tilt towards $\vb{H}_a$ or away from it, in order to lower the energy of the system.
Compared to the equilibrium states of Fig.~\ref{fig:equilibrium}, the angles between the sublattice vectors change, although by a negligible amount.
However, if the applied field is large it could disturb the almost $120^\circ$ relative orientation of the magnetic moments.
Therefore, in this work, we consider relatively small magnetic fields that are sufficient to aid the dynamics (discussed later in Sec.~\ref{sec:SOT}) without disturbing the antiferromagnetic order, viz. $J_E \gg D_M \gg H_0 M_s$.

\vspace{-10pt}
\subsection{Perturbative Analysis}\label{sec:energy_pertur}
\vspace{-5pt}
For a clear understanding of the aforementioned ground states of the monodomain strained-Mn$_3$Sn, we consider the perturbative approach presented in Refs.~[\citen{liu2017anomalous, li2022free, zhang2023current, yoon2023handedness}].
Firstly, we define the sublattice vector as $\vb{m}_i = \qty(\sqrt{1 - u_i^2} \cos{\qty(\varphi_i)}, \sqrt{1 - u_i^2} \sin{\qty(\varphi_i)}, u_i)$, where $\varphi_i$ and $u_i$ are its azimuthal angle and the z-component, respectively.
Secondly, we define the experimentally relevant cluster magnetic octupole moment~\citep{suzuki2017cluster, zhang2023current, yoon2023handedness} as $\vb{m}_\mathrm{oct} = \frac{1}{3} \mathcal{M}_{zx}\qty[R(\frac{2\pi}{3}) \vb{m}_1 + R(-\frac{2\pi}{3}) \vb{m}_2 + \vb{m}_3]$.
Here, $\vb{m}_1$ and $\vb{m_2}$ are rotated by $+2 \pi/3$ and $-2 \pi/3$, respectively, while the y-component of the resultant vector undergoes a mirror operation with respect to x-z plane.~\citep{zhang2023current, yoon2023handedness}
This ensures that the octupole and the sublattice vectors are coplanar, and $\varphi_\mathrm{oct} = -\frac{\varphi_1 + \varphi_2 + \varphi_3}{3}$, where $\varphi_\mathrm{oct}$ is the azimuthal angle of the magnetic octupole \textcolor{black}{moment}.
Thirdly, we define $\varphi_i = - \varphi_\mathrm{oct} - \frac{2\pi i}{3} + \eta_i$, where $\eta_i$ is a small angle ($\eta_i \ll 2\pi/3$) that includes the effect of small deviation from the rigid $120^\circ$ configuration due to both the frustrated bulk structure and the strain.
Here, $\eta_i$ is linearly independent of $\varphi_\mathrm{oct}$ and $\eta_1 + \eta_2 + \eta_3 = 0$.
Finally, we use the perturbative approach, as outlined in the supplementary material, to arrive at an energy landscape, which is a function of $\varphi_\mathrm{oct}$~\citep{liu2017anomalous, li2022free, zhang2023current, yoon2023handedness} and is given as
\begin{flalign}\label{eq:energy_density_oct}
    \begin{split}
        &F\qty(\varphi_\mathrm{oct}) \approx  - \frac{3 A}{2} \cos{\qty(2\varphi_\mathrm{oct})} - \frac{B}{2} \cos{\qty(6\varphi_\mathrm{oct})}  \\
        &- 3 M_s H_0 \qty(C \cos{\qty(\varphi_\mathrm{oct} - \varphi_H)} + D \cos{\qty(\varphi_\mathrm{oct} + \varphi_H)}),
    \end{split}
\end{flalign}
where $A = \frac{2J_E \delta_E K_e}{3\qty(J_E + \sqrt{3} D_M)}$, $B = \frac{\qty(3J_E + 7\sqrt{3}D_M)K_e^3}{9 \qty(J_E + \sqrt{3} D_M)^3}$, $C = \frac{K_e}{3\qty(J_E + \sqrt{3} D_M)}$, and $D = \frac{J_E \delta_E}{3\qty(J_E + \sqrt{3} D_M)}$.
The constant terms are not shown here as they do not affect the ground states solution.

For $H_0 = 0$ in Eq.~(\ref{eq:energy_density_oct}), the $\cos{\qty(2\varphi_\mathrm{oct})}$ term dominates over the $\cos{\qty(6\varphi_\mathrm{oct})}$ term, {if $A \gg B$ and $|A| \gg 3B$ in the case of Mn$_3$Sn with compressive and tensile strains, respectively. 
For the material parameters listed in Table~\ref{tab:mat_params1}, these conditions are equivalent to $\delta_E \gg 0.08 \times 10^{-3}$ and $|\delta_E| \gg 0.24 \times 10^{-3}$, respectively.}
Therefore, {for $|\delta_E| = 10^{-3}$}, 
compressive (tensile) strain leads to two minimum energy equilibrium states of the octupole moment, corresponding to $\varphi_\mathrm{oct} = 0$ and $\pi$ ($\varphi_\mathrm{oct} = \pi/2$ and $3\pi/2$), as shown in Fig.~\ref{fig:equilibrium}.
On the other hand, when a magnetic field is turned on ($H_0 > 0$), the energy of the system changes, and two ground states of the octupole moment, corresponding to the minimum of Eq.~(\ref{eq:energy_density_oct}), are {possible}.
We find that, if $\varphi_H$ is $0^\circ$ or $180^\circ$ ($90^\circ$ or $270^\circ$) for Mn$_3$Sn with compressive (tensile) strain, the possible ground states are same as the initial equilibrium states.
Conversely, if $\varphi_H$ is different from the equilibrium direction, the possible ground states are different from the equilibrium states \textcolor{black}{(Fig.~\ref{fig:gs})}.

\begin{figure}[ht!]
  \centering
  \includegraphics[width = 0.85\columnwidth, clip = true, trim = 0mm 0mm 0mm 0mm]{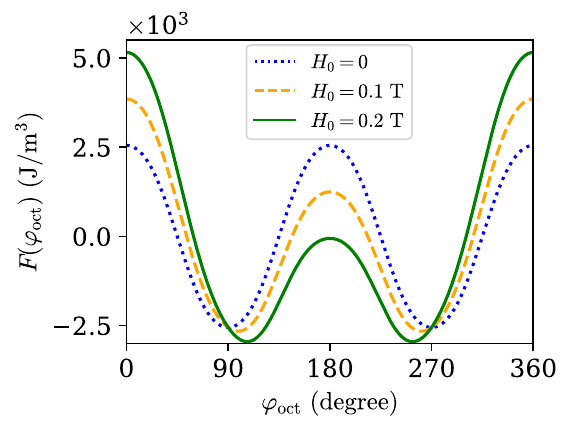}
  \caption{Energy density as a function of the octupole angle of a monodomain Mn$_3$Sn with tensile strain for different applied magnetic field, $H_0$, along the -x-direction ($\varphi_H = 180^\circ$). External magnetic field breaks the symmetry of the system, and therefore, the barrier height reduces at $\varphi_\mathrm{oct} = \pi$ but increases at $\varphi_\mathrm{oct} = 0$, as $H_0$ increases.} 
  \label{fig:energy}
\end{figure}

In the special case of $\vb{H}_a$ perpendicular to the equilibrium direction, degenerate ground states with energies lower than that of the equilibrium states are obtained.
This is \textcolor{black}{depicted} in Fig.~\ref{fig:energy}, where an external magnetic field is applied in the negative x-direction ($\varphi_H = 180^\circ$) to Mn$_3$Sn with tensile strain.
As the strength of the magnetic field increases, the energy of the ground states decrease and they move away from the equilibrium states of $90^\circ$ and $270^\circ$, towards $180^\circ$. 
In addition, the energy barrier separating the two states reduces at $180^\circ$ but increases at $360^\circ$. 
The  two ground states exist if $H_0 \lesssim \frac{2|A+3B|}{M_s \qty(C+D)}$. 
\textcolor{black}{Within this limit, $\norm{\vb{m}}$ was found to increase with both $\delta_E$ and $H_0$ (see supplementary material).}
For higher $H_0$, $\varphi_\mathrm{oct} = \pi$ becomes a minimum too \textcolor{black}{and $\norm{\vb{m}}$ decreases with $\delta_E$ at fixed $H_0$ (supplementary material).}  
In the case of Mn$_3$Sn with compressive strain and $\vb{H}_a$ perpendicular to the equilibrium states, two ground states exist if $H_0 \lesssim \frac{2(A+3B)}{M_s \qty(C-D)}$. 
These limits suggest that the maximum value of $H_0$, which could be applied while maintaining two ground states, increases with $\delta_E$.
This is because larger $\delta_E$ leads to higher intrinsic energy barrier, which is given as $|3A + B|$, between the two equilibrium states.

\vspace{-10pt}
\section{SOT-driven dynamics}\label{sec:SOT}
\begin{figure}[ht!]
  \centering
  \includegraphics[width = 0.85\columnwidth, clip = true, trim = 0mm 0mm 0mm 0mm]{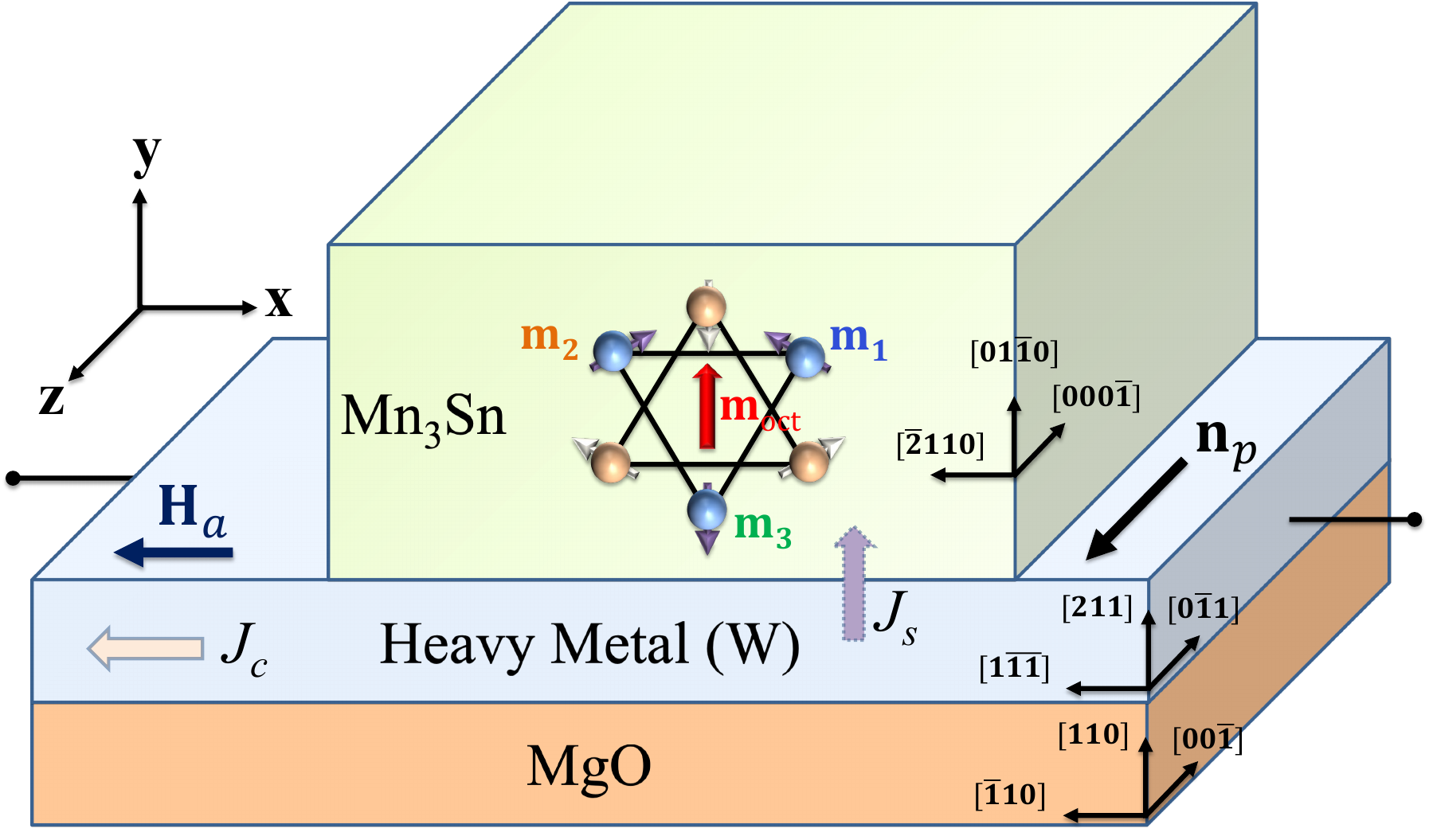}
  \caption{Spin-orbit torque device setup for manipulating the magnetic state in Mn$_3$Sn. 
  The MgO substrate leads to tensile strain in the x-direction, and hence the PMA in Mn$_3$Sn.
  Spin-Hall effect in the HM generates SOT when a charge current flows into it.
  $J_c$ and $J_s$ are the charge current density and the spin current density, respectively.
  $\vb{H}_a$ is the external magnetic field which is applied to aid the deterministic switching of the magnetic octupole \textcolor{black}{moment}, $\vb{m}_\mathrm{oct}$, in strained-Mn$_3$Sn.
  } 
  \label{fig:device}
\end{figure}

\vspace{-5pt}
To investigate the dynamics of Mn$_3$Sn under the effect of spin current, we consider the spin-Hall effect (SHE) setup shown in Fig.~\ref{fig:device}. This setup resembles the experimental designs from Refs.~[\citen{higo2022perpendicular, yoon2023handedness}], where Mn$_3$Sn grown epitaxially on a $(110)[001]$ MgO substrate exhibits uniaxial tensile strain in the x-direction {resulting in} a PMA energy landscape {for the magnetic octupole \textcolor{black}{moment}}.~\citep{higo2022perpendicular, yoon2023handedness}
Hereafter, we only focus on the dynamics of single-domain Mn$_3$Sn with tensile strain while the discussion on the dynamics of Mn$_3$Sn with compressive strain and no strain is relegated to supplementary material. 
In our convention, as mentioned previously, the Kagome plane of Mn$_3$Sn is assumed to coincide with the x-y plane while the z-axis coincides with $[0001]$ direction.
Charge-to-spin conversion in the HM, due to the flow of charge current density, $J_c$, leads to the generation of a spin current density, $J_s$, polarized along $\vb{n}_p$, which is assumed to coincide with z-axis. 
Previous works have shown that the current required to induce dynamics in this setup, with $\vb{n}_p$ perpendicular to the Kagome plane, is significantly smaller than that required in the case where $\vb{n}_p$ is parallel to the Kagome plane.~\citep{takeuchi2021chiral, higo2022perpendicular, yoon2023handedness}
Finally, the external field $\vb{H}_a$ is assumed to be applied in the negative x-direction, or $\varphi_H = 180^\circ$.

For each sublattice of Mn$_3$Sn, the magnetization dynamics is governed by the classical 
Landau-Lifshitz-Gilbert (LLG) equation, which is a statement of the conservation of angular momentum. The LLG equations for the three sub-lattices are coupled via the exchange interactions.~\citep{yamane2019dynamics, shukla2022spin}
For sublattice $i$, the LLG equation is given as~\citep{mayergoyz2009nonlinear}
\begin{equation}\label{eq:sLLGS}
    \begin{split}
       \dot{\vb{m}}_i &= - \gamma \qty(\vb{m}_i \cp \vb{H}_i^\mathrm{eff}) + \alpha \qty(\vb{m}_i \cp \dot{\vb{m}}_i) \\  
        & - \frac{\hbar}{2e}\frac{\gamma J_{s}}{M_{s} d_{a}}\vb{m}_i \cp \qty(\vb{m}_i \cp \vb{n}_{p}),
    \end{split}
\end{equation}
where $\dot{\vb{m}}_{i} = \pdv{\vb{m}_i}{t}$, $t$ is time in seconds, $\vb{H}_{i}^{\mathrm{eff}}$ is the effective magnetic field experienced by $\vb{m}_i$, $\alpha$ is the Gilbert damping parameter for Mn$_3$Sn, and $d_{a}$ is the thickness of the AFM layer. Other parameters in this equation, viz. $\hbar = 1.054561 \times 10^{-34}~\mathrm{J~s}$, $e = 1.6 \times 10^{-19} ~\mathrm{C}$, and $\gamma = 17.6 \cp 10^{10}~\mathrm{T^{-1}~s^{-1}}$ are the reduced Planck's constant, the elementary charge of an electron, and the gyromagnetic ratio, respectively.
The spin current density depends on the input charge current density and the spin-Hall angle of the HM, $\theta_\mathrm{SH}$, as $J_s = \theta_\mathrm{SH} J_c$. The spin-Hall angle is associated with the efficiency of the SOT effect.
Here, we consider the HM to be W since it has a large $\theta_\mathrm{SH}$~\citep{tsai2020electrical}.

\begin{figure}[ht!]
  \centering
  \includegraphics[width = \columnwidth, clip = true, trim = 0mm 0mm 0mm 0mm]{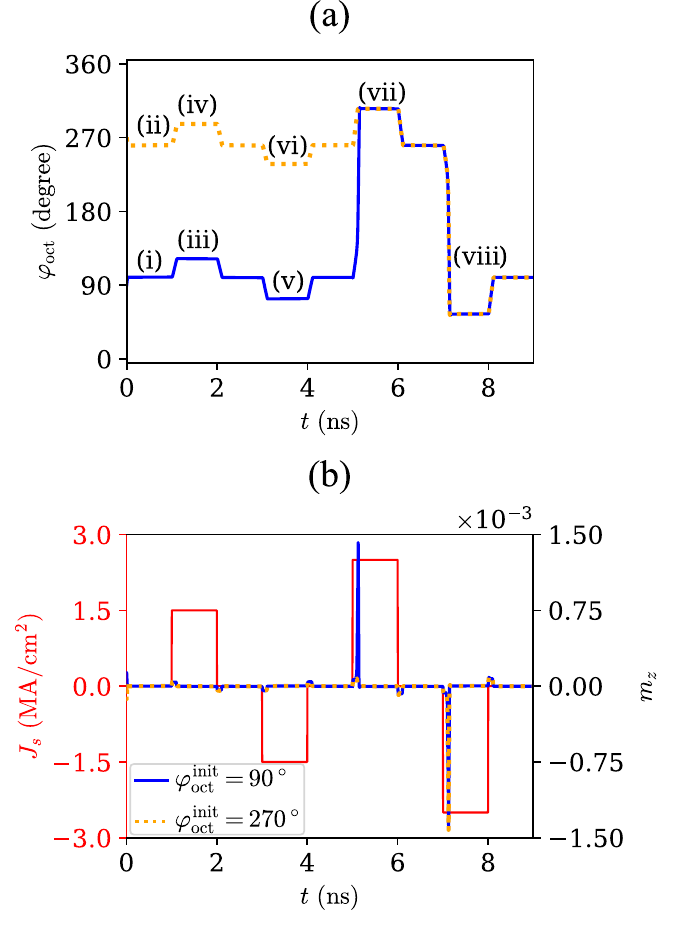}
  \caption{{Steady-state response of (a) the magnetic octupole moment and (b) the out-of-(Kagome)-plane component of the average magnetization, $m_z$, as a function of time under the effect of a current pulse, which is turned on at $t = 1~\mathrm{ns}$ and $t = 3~\mathrm{ns}$ to $|J_s| = 1.5~\mathrm{MA/cm^2}$ while it is increased to $|J_s| = 2.5~\mathrm{MA/cm^2}$ at $t = 5~\mathrm{ns}$ and $t = 7~\mathrm{ns}$. Otherwise the current is turned off. 
  (a) (i) and (ii): Possible ground states for $H_0 = 0.1~\mathrm{T}$ in the negative x-direction and $J_s = 0$. 
  (iii) and (iv): $J_s = 1.5~\mathrm{MA/cm^2}$. No switching. Stationary states at angles larger than (i) and (ii), respectively.
  (v) and (vi): $J_s = -1.5~\mathrm{MA/cm^2}$. No switching. Stationary states at angles smaller than (i) and (ii), respectively. 
  (vii): $J_s = 2.5~\mathrm{MA/cm^2}$. Deterministic switching of (i). No switching for (ii). Stationary states at angles larger than (ii).
  (viii): $J_s = -2.5~\mathrm{MA/cm^2}$. Deterministic switching of (ii). Stationary states at angles smaller than (i). 
  (b) $|m_z|$ is zero in the steady state but increases during the switching process. The change in $|m_z|$ is negligible for the case of no switching.}  
  } 
  \label{fig:phi_oct_ns_s}
\end{figure}

\begin{figure}[ht!]
  \centering
  \includegraphics[width = 0.85\columnwidth, clip = true, trim = 0mm 0mm 0mm 0mm]{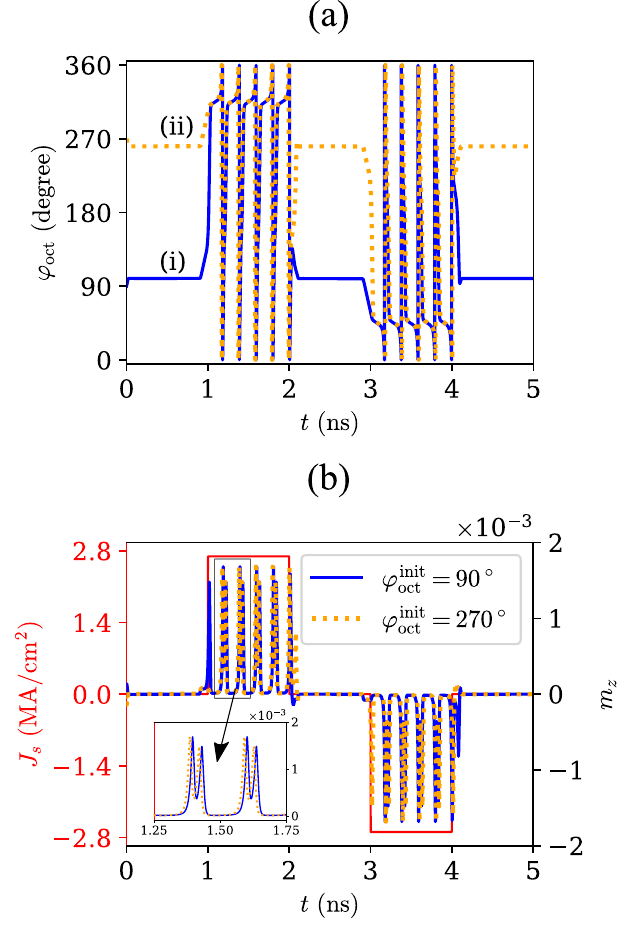}
  \caption{{Steady-state oscillation dynamics of (a) the magnetic octupole moment and (b) the out-of-(Kagome)-plane component of the average magnetization, $m_z$, as a function of time under the effect of a current pulse.
  The pulse is turned on at $t = 1~\mathrm{ns}$ ($t = 3~\mathrm{ns}$) to $J_s = 2.69~\mathrm{MA/cm^2}$ ($J_s = -2.69~\mathrm{MA/cm^2}$) and turned off at $t = 2~\mathrm{ns}$ ($t = 4~\mathrm{ns}$). 
  (a) (i) and (ii): Possible ground states for $H_0 = 0.1~\mathrm{T}$ in the negative x-direction. 
  Oscillation dynamics for $|J_s| = 2.69~\mathrm{MA/cm^2}$: the octupole moment in both (i) and (ii) oscillate at a frequency of about $2.5~\mathrm{GHz}$ in a direction decided by the direction of the input current. When the current is turned off, the octupole moment probabilistically settles into either of (i) or (ii). 
  (b) Non-zero $m_z$ facilitates high frequency chiral oscillations in the steady state due to the strong exchange interaction between out-of-(Kagome)-plane components of the sublattice vectors.}  
  } 
  \label{fig:phi_oct_osc1}
\end{figure}

The effective magnetic field for sublattice $i$ can be obtained by using Eq.~(\ref{eq:energy_density}) as
\begin{flalign}\label{eq:m_i_field}
    \begin{split}
        \vb{H}_{i}^{\mathrm{eff}}  &= -\frac{1}{M_{s}}\pdv{F}{\vb{m}_{i}} = -\frac{J_E}{M_{s}} \qty(\vb{m}_{j} + \vb{m}_k)  \\
        &+\frac{D_M \vb{z} \cp \qty(\vb{m}_{j} - \vb{m}_{k})}{M_{s}} + \frac{2K_{e}}{M_{s}} \qty(\vb{m}_{i} \vdot \vb{u}_{e, i}) \vb{u}_{e, i} + \vb{H}_{a},
    \end{split}
\end{flalign}
where $(i, j, k) = (1, 2, 3), (2, 3, 1),$ or $(3, 1, 2)$, respectively. 
Equations~(\ref{eq:sLLGS}) and~(\ref{eq:m_i_field}) are then solved simultaneously, for a range of $H_0 (>0)$ and $J_s$, for both $\varphi_\mathrm{oct}^\mathrm{init} = \pi/2$ {and $\varphi_\mathrm{oct}^\mathrm{init} = 3\pi/2$} as the initial states. The steady-state response of the magnetic order of Mn$_3$Sn is found to be dependent on the initial ground states, magnitude of $H_0$, and the direction and magnitude of the input current.
These differences in the steady state behavior are shown in Figs.~\ref{fig:phi_oct_ns_s} and~\ref{fig:phi_oct_osc1} for $H_0 = 0.1~\mathrm{T}$ and $\varphi_H = 180^\circ$.

\textcolor{black}{Figure~\ref{fig:phi_oct_ns_s} shows the time dynamics of the magnetic octupole moment and that of the out-of-(Kagome)-plane component of the average magnetization, $m_z$, for the current pulse shown in Fig.~\ref{fig:phi_oct_ns_s}(b).
It can be observed that for $t < 1~\mathrm{ns}$, where $J_s = 0$, the magnetic octupole moment evolves to ground states (i) and (ii), for the equilibrium states at $\varphi_\mathrm{oct}^\mathrm{init} = \pi/2$ and $\varphi_\mathrm{oct}^\mathrm{init} = 3\pi/2$, respectively.
When $J_s$ is increased to $1.5~\mathrm{MA/cm^2}$ at $t = 1~\mathrm{ns}$, (i) and (ii) evolve to stationary steady-states (iii) and (iv), respectively.
Although the $\varphi_\mathrm{oct}$ corresponding to both (iii) and (iv) are larger than those for (i) and (ii), they are still near the initial states, that is, $1.5~\mathrm{MA/cm^2}$ is not large enough to induce deterministic switching.
Stationary steady-states near the initial ground states are also observed when $J_s$ is decreased to $-1.5~\mathrm{MA/cm^2}$ at $t = 3~\mathrm{ns}$: (i) and (ii) evolve to (v) and (vi), respectively. In this case, the $\varphi_\mathrm{oct}$ corresponding to the stationary states are smaller than those of the ground states.
States (iii) and (v) return to the ground state (i) while (iv) and (vi) return to the ground state (ii), when the current is turned off.
The final steady-states, therefore, depend on both the magnitude and the direction of the input current.} 
As shown in Fig.~\ref{fig:phi_oct_ns_s}(b), $m_z$ is zero in the steady state. It changes negligibly, in the direction of the change in $J_s$, when the current is turned on or off.

\textcolor{black}{It can be further observed from Fig.~\ref{fig:phi_oct_ns_s}(a) that the magnetic octupole moment in both the ground states (i) and (ii) evolve to the stationary steady-state (vii), when $J_s$ is increased to $2.5~\mathrm{MA/cm^2}$ at $t = 5~\mathrm{ns}$.
On the one hand, this dynamics corresponds to the deterministic switching of the magnetic octupole moment in (i). 
On the other hand, for the magnetic octupole moment in (ii), (vii) is just a stationary steady-state with $\varphi_\mathrm{oct}$ greater than that of (ii).}
\textcolor{black}{However, when the current direction is reversed by lowering $J_s$ to $-2.5~\mathrm{MA/cm^2}$ at $t = 7~\mathrm{ns}$, the magnetic octupole moment in (ii) switches deterministically to (viii), which is near (i) but has a smaller corresponding $\varphi_\mathrm{oct}$.}
In this case, the \textcolor{black}{magnetic octupole moment} in (i) will not switch, but move to (viii).
As shown in Fig.~\ref{fig:phi_oct_ns_s}(b) at $t = 5~\mathrm{ns}$ and $t = 7~\mathrm{ns}$, deterministic switching is accompanied by a large spike in $m_z$. The direction of change in $m_z$ depends on the direction of the input current---positive (negative) $J_s$ leads to positive (negative) $m_z$.

\textcolor{black}{Further increasing $|J_s|$ to $2.69~\mathrm{MA/cm^2}$ results in chiral oscillations for the magnetic octupole moment in both the ground states (i) and (ii), as shown in Fig.~\ref{fig:phi_oct_osc1}(a).
For positive $J_s$, the magnetic octupole \textcolor{black}{moment} in (i) 
deterministically switches to (ii) in the first step; therefore, the phase of (i) lags that of (ii). 
On the contrary, for negative $J_s$, the magnetic octupole \textcolor{black}{moment} in (ii) deterministically switch to (i) in the first step; therefore, the phase of (ii) lags that of (i).}
The oscillation dynamics of the magnetic octupole \textcolor{black}{moment} is accompanied by large $m_z$, as shown in Fig.~\ref{fig:phi_oct_osc1}(b). Similar to the case of deterministic switching dynamics, the direction of $m_z$ depends on the direction of current---positive (negative) $J_s$ leads to positive (negative) $m_z$.
\textcolor{black}{However, unlike the case of deterministic switching, $m_z$ shows two spikes per oscillation (inset of Fig.~\ref{fig:phi_oct_osc1}(b)).}

\begin{figure}[ht!]
  \centering
  \includegraphics[width = 0.85\columnwidth, clip = true, trim = 0mm 0mm 0mm 0mm]{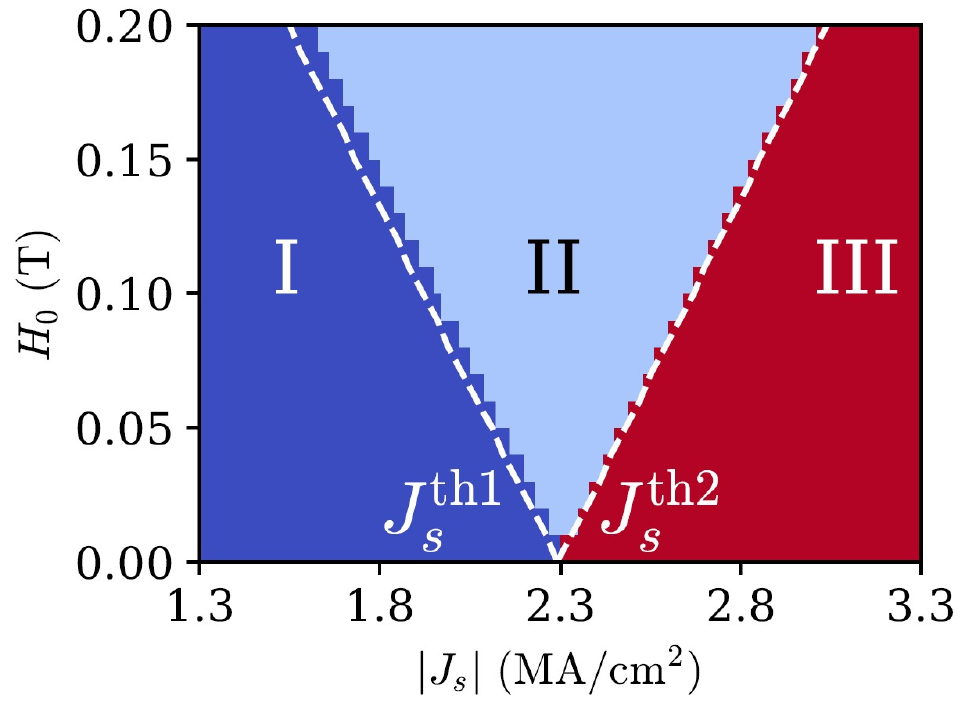}
  \caption{Final steady-state as a function of the magnitude of the input current, $|J_s|$, for different applied magnetic fields $H_0$. I, II, and III represent regions of no switching, switching, and \textcolor{black}{chiral oscillation}, respectively. The dashed white lines represent the two threshold currents, $J_s^\mathrm{th1}$ and $J_s^\mathrm{th2}$ obtained from numerical solution of Eq.~(\ref{eq:stationary_state}). 
  {This phase diagram is applicable if $\varphi_\mathrm{oct}^\mathrm{init} = \pi/2$ ($\varphi_\mathrm{oct}^\mathrm{init} = 3\pi/2$) and $J_s > 0$ ($J_s < 0$). 
  For $\varphi_\mathrm{oct}^\mathrm{init} = 3\pi/2$ ($\varphi_\mathrm{oct}^\mathrm{init} = \pi/2$) and $J_s > 0$ ($J_s < 0$), region III would still represents chiral oscillation, however, the regions encompassing I and II would both correspond to no switching.}} 
  \label{fig:final_state}
\end{figure}

Detailed numerical simulations revealed that for $\varphi_\mathrm{oct}^\mathrm{init} = \pi/2$ ($\varphi_\mathrm{oct}^\mathrm{init} = 3\pi/2$) and $J_s > 0$ ($J_s < 0$), the final steady-state of $\vb{m}_\mathrm{oct}$ depended on the magnitude of $J_s$ with respect to two threshold currents---$J_s^\mathrm{th1}$ and $J_s^\mathrm{th2}$, where $J_s^\mathrm{th1} < J_s^\mathrm{th2}$.
As summarized in Fig.~\ref{fig:final_state},
if the injected current density is smaller than the lower threshold current, that is $|J_s| < J_s^\mathrm{th1}$, the ground state of the AFM evolves to a non-equilibrium stationary steady-state in the initial energy well. 
\textcolor{black}{The case of $|J_s| = 1.5~\mathrm{MA/cm^2}$ and $H_0 = 0.1~\mathrm{T}$, shown in Fig.~\ref{fig:phi_oct_ns_s}(a), belongs to this regime.}
On the other hand, if $J_s^\mathrm{th1} < |J_s| < J_s^\mathrm{th2}$ the \textcolor{black}{magnetic octupole moment} deterministically switches to a non-equilibrium stationary steady-state in the other energy well \textcolor{black}{($|J_s| = 2.5~\mathrm{MA/cm^2}$ and $H_0 = 0.1~\mathrm{T}$ in Fig.~\ref{fig:phi_oct_ns_s}(a))}.
Finally, when $J_s^\mathrm{th2} < |J_s|$ the magnetic octupole moment exhibits steady-state chiral oscillations \textcolor{black}{($|J_s| = 2.69~\mathrm{MA/cm^2}$ and $H_0 = 0.1~\mathrm{T}$ in Fig.~\ref{fig:phi_oct_osc1}(a))}, whose frequency could be tuned from the 100s of MHz to the {10s of GHz} range by varying $|J_s|$.
The three regimes of operation are marked as I, II, and III for no switching, deterministic switching, and chiral-rotation, respectively.
The overlaid dashed white lines represent $J_s^\mathrm{th1}$ and $J_s^\mathrm{th2}$. It can be observed that $J_s^\mathrm{th1}$ decreases with an increase in $H_0$ while $J_s^\mathrm{th2}$ increases with $H_0$.
As a result, the range of input currents where the system exhibits deterministic switching increases with $H_0$.
For $\varphi_\mathrm{oct}^\mathrm{init} = 3\pi/2$ ($\varphi_\mathrm{oct}^\mathrm{init} = \pi/2$) and $J_s > 0$ ($J_s < 0$), the \textcolor{black}{magnetic octupole moment} displays a stationary steady-state in the initial energy well, if $|J_s| < J_s^\mathrm{th2}$ \textcolor{black}{(Fig.~\ref{fig:phi_oct_ns_s}(a))}, while it shows chiral oscillations for $J_s^\mathrm{th2} < |J_s|$ \textcolor{black}{(Fig.~\ref{fig:phi_oct_osc1}(a))}.
In the limiting case of $H_0 = 0$, $J_s^\mathrm{th1} = J_s^\mathrm{th2}$, and no deterministic switching of the \textcolor{black}{magnetic octupole moment} is observed. Instead, the \textcolor{black}{magnetic octupole moment} displays either a non-equilibrium stationary state in the initial energy well or chiral oscillations. 
If the current is turned off during the oscillation, the \textcolor{black}{magnetic octupole moment} probabilistically switches to either of the energy wells.~\cite{takeuchi2021chiral, dasgupta2022tuning, shukla2023order}

\vspace{-10pt}
\subsection{Stationary State and Threshold Current}
\vspace{-5pt}
To explore the dependence of the dynamics on the intrinsic energy scale of the system, obtain analytic expressions for the two threshold currents as a function of the applied magnetic field and material parameters, and establish scaling laws related to switching and chiral oscillations, we evaluate the rate of change of the average magnetization, $\dot{\vb{m}} = \frac{\qty(\dot{\vb{m}}_1 + \dot{\vb{m}}_2 + \dot{\vb{m}}_3)}{3}$, as
\begin{flalign}\label{eq:m_dot1}
    \begin{split}
        \dot{\vb{m}} &= \frac{1}{3}\sum_{i=1}^{3} \bigg(-\gamma \qty(\vb{m}_{i} \cp \vb{H}_{i}^\mathrm{eff}) + \alpha \qty(\vb{m}_i \cp \dot{\vb{m}}_i) \Bigg. \\
        &\left.- \frac{\hbar}{2e}\frac{\gamma J_{s}}{M_{s} d_{a}}\qty(\vb{m}_i \cp \qty(\vb{m}_i \cp \vb{z})) \right),
    \end{split}
\end{flalign}
where $\vb{m}_i \cp \vb{H}_{i}^\mathrm{eff} =  \frac{1}{3\mu_0 M_s} \pdv{F}{\varphi_\mathrm{oct}} \vb{z}$ while $\pdv{F}{\varphi_\mathrm{oct}}$ is obtained from Eq.~(\ref{eq:energy_density_oct}) with $\varphi_H = \pi$. 

In the stationary states, \textcolor{black}{irrespective of $\varphi_\mathrm{oct}^\mathrm{init}$}, the net torque on the \textcolor{black}{magnetic octupole moment is zero since the spin-orbit torque generated by the input current is balanced by the torque due to the internal and external magnetic fields.}
\textcolor{black}{Consequently}, we set the time derivatives ($\dot{\vb{m}}$ and $\dot{\vb{m}}_i$) in Eq.~(\ref{eq:m_dot1}) to zero. 
Our numerical simulations revealed that the z-component of all the sublattice vectors were zero in the stationary steady-state.
So, we set $m_{i, z} = u_i = 0$ in Eq.~(\ref{eq:m_dot1}) to arrive at the torque balance equation:
\begin{flalign}\label{eq:stationary_state}
    \begin{split}
         A\sin{\qty(2\varphi_\mathrm{oct})} + B \sin{\qty(6\varphi_\mathrm{oct})} - M_s H_0 \qty(C + D) \sin{\qty(\varphi_\mathrm{oct})} \\
        = \frac{\hbar}{2e}\frac{J_{s}}{d_{a}}. 
    \end{split}
\end{flalign}

\begin{figure}[ht!]
  \centering
  \includegraphics[width = 0.85\columnwidth, clip = true, trim = 0mm 0mm 0mm 0mm]{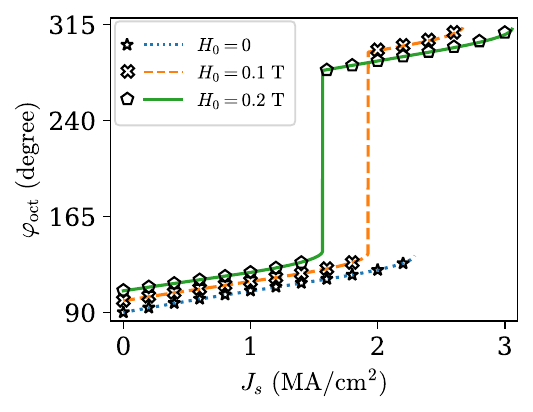}
  \caption{Stationary steady-states as a function of the applied spin current, $J_s$, for a thin film of Mn$_3$Sn under tensile strain. Numerical result from the solution of Eq.~(\ref{eq:sLLGS}) (symbols) agree very well with the results obtained from Eq.~(\ref{eq:stationary_state}) (lines)} 
  \label{fig:stationary}
\end{figure}

For $\varphi_\mathrm{oct}^\mathrm{init} = \pi/2$ and $0 \leq J_s < J_s^\mathrm{th1}$, the solution to Eq.~(\ref{eq:stationary_state}) should lead to one stationary solution with $\mathrm{GS}_1 \leq \varphi_\mathrm{oct} < \pi $, where $\mathrm{GS}_1 \in (\pi/2, \pi)$ is the smaller of the two minima of Eq.~(\ref{eq:energy_density_oct}). 
On the other hand, for $J_s^\mathrm{th1} \leq J_s < J_s^\mathrm{th2}$, the \textcolor{black}{magnetic octupole moment} should switch to a stationary state in the other energy well, and the solution of Eq.~(\ref{eq:stationary_state}) should lead to $\mathrm{GS}_2 < \varphi_\mathrm{oct} < 2\pi$. 
Here, $\mathrm{GS}_2 \in (\pi, 3\pi/2)$ is the larger of the two minima of Eq.~(\ref{eq:energy_density_oct}). 
Indeed the same can be observed from Fig.~\ref{fig:stationary}, where the numerical solutions of the coupled LLG equations (symbols) fit the solutions from Eq.~(\ref{eq:stationary_state}) (lines) very well in both the energy wells, for three different values of $H_0$.
Consequently, $J_s^\mathrm{th1}$ is the minimum current for which Eq.~(\ref{eq:stationary_state}) has no solution in $(\mathrm{GS}_1, \pi)$, but has a solution in $(\mathrm{GS}_2, 2\pi)$.
On the other hand, $J_s^\mathrm{th2}$ is the minimum current for which Eq.~(\ref{eq:stationary_state}) has no solutions.
The numerical solution of the threshold currents for different $H_0$, obtained from Eq.~(\ref{eq:stationary_state}), is shown by the dashed white lines overlaid on Fig.~\ref{fig:final_state}.
It can be observed that the solutions from Eq.~(\ref{eq:stationary_state}) match the results from Eq.~(\ref{eq:sLLGS}) very well.
If $\varphi_\mathrm{oct}^\mathrm{init} = 3\pi/2$ and $0 \leq J_s < J_s^\mathrm{th2}$, the solution of Eq.~(\ref{eq:stationary_state}) would lead to $\mathrm{GS}_2 \leq \varphi_\mathrm{oct} < 2\pi$, as is shown in Fig.~\ref{fig:stationary} for $J_s^\mathrm{th1} \leq J_s < J_s^\mathrm{th2}$. 
\textcolor{black}{Although not shown here, reversing the direction of current ($J_s < 0$) with $0 \leq |J_s| < J_s^\mathrm{th2}$ leads to stationary steady-states in $(0, \mathrm{GS}_1]$ for $\varphi_\mathrm{oct}^\mathrm{init} = \pi/2$ while $\varphi_\mathrm{oct}^\mathrm{init} = 3\pi/2$ exhibits a stationary state in $(\pi, \mathrm{GS}_2]$, if $0 \leq |J_s| < J_s^\mathrm{th1}$, and in $(0, \mathrm{GS}_1)$, if $J_s^\mathrm{th1} \leq |J_s| < J_s^\mathrm{th2}$.}

\begin{figure}[ht!]
    \centering
    \includegraphics[width = 0.85\columnwidth, clip = true, trim = 0mm 0mm 0mm 0mm]{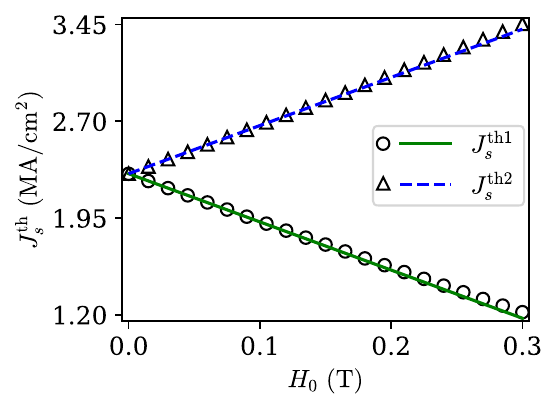}
    \caption{Comparison of the analytic expressions of threshold currents (Eq.~(\ref{eq:threshold})), shown by lines, against the numerical values obtained from the solution of Eq.~(\ref{eq:stationary_state}), represented by symbols. The numerical and analytic values show excellent agreement for the values of $H_0$ considered here.}
    \label{fig:threshold_current}
\end{figure}

An exact expression of either of the two threshold currents is cumbersome to obtain, however,
in the limit of small magnetic fields that do not disturb the two-fold degeneracy of PMA Mn$_3$Sn, they can be approximated as
\begin{subequations}\label{eq:threshold}
    \begin{flalign}
        J_s^\mathrm{th1} &= d_a\frac{2e}{\hbar}\qty(-A + B - \frac{M_s H_0}{\sqrt{2}}\qty(C + D)), \\
        J_s^\mathrm{th2} &= d_a\frac{2e}{\hbar}\qty(-A + B + \frac{M_s H_0}{\sqrt{2}}\qty(C + D)).
    \end{flalign}
\end{subequations}
In the absence of an external magnetic field, the threshold current is the minimum current that provides just enough SOT to overcome the maximum torque due to the effective in-plane anisotropy.
\textcolor{black}{This maximum occurs at} $\varphi_\mathrm{oct} = 45^\circ$, $135^\circ$, $225^\circ$, and $315^\circ$ since they lead to $\sin{\qty(2\varphi_\mathrm{oct})} = \pm 1$ and $\sin{\qty(6\varphi_\mathrm{oct})} = \mp 1$.
\textcolor{black}{For non-zero external magnetic field, first, we consider the effect of the in-plane anisotropy to be dominant while that of $H_0$ to be small. 
We then evaluate Eq.~(\ref{eq:stationary_state}) at $\varphi_\mathrm{oct} = 135^\circ$ and $\varphi_\mathrm{oct} = 315^\circ$ to obtain $J_s^\mathrm{th1}$ and $J_s^\mathrm{th2}$, respectively.}
Figure~\ref{fig:threshold_current} compares the analytic expressions of Eq.~(\ref{eq:threshold}) (lines) against the values of the threshold currents obtained from the solution of Eq.~(\ref{eq:stationary_state}) (symbols), for different values of $H_0$. It can be observed that the analytic results match very well against the numerical values. 
\textcolor{black}{Since the torque due to $H_0$ acts against (along) the torque due to the effective in-plane anisotropy at $\varphi_\mathrm{oct} = 135^\circ$ ($\varphi_\mathrm{oct} = 315^\circ$), larger $H_0$ reduces (increases) $J_s^\mathrm{th1}$ ($J_s^\mathrm{th2}$).}
Although the error between the numerical and the analytic values of the two threshold currents increases with an increase in $H_0$, it is still smaller than $5 \%$ even for $H_0 = 0.3~\mathrm{T}$.
This linear dependence of the threshold currents on the external field is similar to that in the case of a PMA ferromagnet driven by a SOT.~\citep{lee2013threshold}

\vspace{-5pt}
\subsection{Deterministic Switching Dynamics}
\vspace{-5pt}
For $H_0 > 0$ and $J_s^\mathrm{th1} \leq |J_s| < J_s^\mathrm{th2}$, \textcolor{black}{the time derivatives in Eq.~(\ref{eq:m_dot1}) change to non-zero values ($|\dot{\vb{m}}| > 0$ and $|\dot{\vb{m}}_i| > 0$), if $\varphi_\mathrm{oct}^\mathrm{int} = 90^\circ$ ($\varphi_\mathrm{oct}^\mathrm{int} = 270^\circ$) and $J_s > 0$ ($J_s < 0$). 
Here, the external magnetic field assists the SOT in overcoming the maximum torque due to the internal magnetic fields.
Consequently, the magnetic octupole moment moves away from its initial stable state and switches over the energy barrier at $\varphi_\mathrm{oct} = 180^\circ$ to the other energy well, while both $|m_z|$ and $|u_i|$ increase to non-zero values. 
In the second energy well, initially, the torque due to the magnetic fields and the SOT act in the same direction, leading to further increase in $|m_z|$ until $\dot{\vb{m}}$ decreases to zero due to the effect of the intrinsic damping. 
Subsequently, $|m_z|$ decreases to zero while the octupole moment slows down and reaches a stationary state. The SOT cannot overcome the torques due to the magnetic fields anymore since the external field aids the internal fields in the second energy well.} 
If, however, the input current is reversed, such that $J_s^\mathrm{th1} \leq |J_s| < J_s^\mathrm{th2}$, for the same $H_0$, the \textcolor{black}{magnetic octupole moment goes back to the initial energy well, by crossing the barrier at $\varphi_\mathrm{oct} = 180^\circ$ as the external field assists the SOT.} 
This bidirectional switching behavior was clearly demonstrated in Fig.~\ref{fig:phi_oct_ns_s} \textcolor{black}{for $|J_s| = 2.5~\mathrm{MA/cm^2}$ and $H_0 = 0.1~\mathrm{T}$.}
\textcolor{black}{Instead of reversing $J_s$, if $\vb{H}_a$ was reversed to the positive x-direction, the deterministic switching dynamics would have proceeded by crossing the barrier at $\varphi_\mathrm{oct} = 360^\circ$, provided that $J_s^\mathrm{th1} \leq |J_s| < J_s^\mathrm{th2}$.}

\begin{figure}[ht!]
  \centering
  \includegraphics[width = 0.85\columnwidth, clip = true, trim = 0mm 0mm 0mm 0mm]{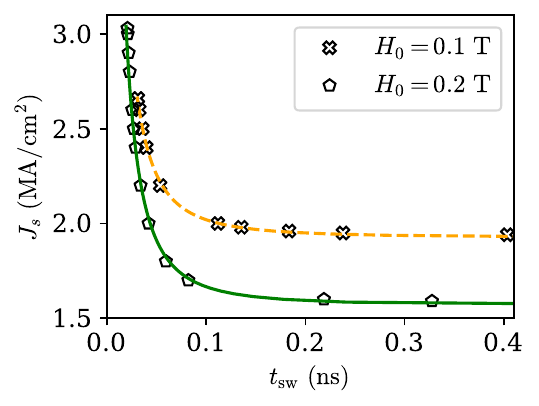}
  \caption{Switching time, $t_\mathrm{sw}$, as a function of the applied spin current, $J_s$, and external magnetic field, $H_0$, for a thin film of Mn$_3$Sn under tensile strain. In each case, the applied current is above $J_s^\mathrm{th1}$ and below $J_s^\mathrm{th2}$. $t_\mathrm{sw}$ obtained from the solution of Eq.~(\ref{eq:tsw}) (lines) fits the data obtained from the solution of Eq.~(\ref{eq:sLLGS}) (symbols) very well.} 
  \label{fig:switching_time}
\end{figure}

\textcolor{black}{The SOT-driven bidirectional deterministic switching dynamics in PMA Mn$_3$Sn could be useful for building next-generation antiferromagnetic memory devices.
In this regard, the switching time, $t_\mathrm{sw}$, as a function of the input current is an important metric.
Figure~\ref{fig:switching_time} shows the $t_\mathrm{sw}$ as function of $J_s$ for two different magnetic fields.} 
Here, $t_\mathrm{sw}$ is defined as the time taken by the magnetic octupole \textcolor{black}{moment} in the ground state $\varphi_\mathrm{oct}^\mathrm{init} = \pi/2$ to go from $\varphi_\mathrm{oct} = \pi/2$ to $\varphi_\mathrm{oct} = \pi$.
This is the minimum duration of an input current pulse that can induce deterministic switching. Such a pulse ensures that the \textcolor{black}{magnetic octupole moment} reaches the top of the energy barrier. Thereafter, the current pulse is turned off and \textcolor{black}{the torques due to the magnetic fields} assist in switching to the other energy well.
It can be observed from Fig.~\ref{fig:switching_time} that $t_\mathrm{sw}$ decreases with an increase in either $J_s$ or $H_0$. 
For a fixed $H_0$, $t_\mathrm{sw}$ decreases with an increase in $J_s$ since a higher input current leads to \textcolor{black}{a larger SOT on the magnetic octupole moment.} 
On the other hand, at a fixed $J_s$, $t_\mathrm{sw}$ decreases with an increase in $H_0$ since it lowers the energy barrier at $\varphi_\mathrm{oct} = 180^\circ$, as shown in Fig.~\ref{fig:energy}.

Our numerical simulations showed that both $u_i$ and $m_z$ were \textcolor{black}{relatively small as the magnetic octupole moved from $\varphi_\mathrm{oct} = \pi/2$ to $\varphi_\mathrm{oct} = \pi$}.
Therefore, the switching time was obtained from the z-component of Eq.~(\ref{eq:m_dot1}) as 
\begin{flalign}\label{eq:tsw}
    \begin{split}
        &t_\mathrm{sw} = \frac{\alpha M_s}{\gamma} \frac{2e}{\hbar} \frac{d_a}{J_s}\int_{\pi/2}^{\pi} \\
        &\times \frac{d \varphi_\mathrm{oct}^{'}}{1 - \frac{\qty(A \sin{\qty(2\varphi_\mathrm{oct}^{'})} + B \sin{\qty(6\varphi_\mathrm{oct}^{'})} - M_s H_0 \qty(C + D) \sin{\qty(\varphi_\mathrm{oct}^{'})})}{\frac{\hbar}{2 e}\frac{J_s}{d_a}}},
    \end{split}
\end{flalign}
where we neglected the rate of change of $m_z$ since $\pdv{m_z}{t} \ll \alpha \pdv{\varphi_\mathrm{oct}}{t}$. It can be observed from Fig.~\ref{fig:switching_time} that the switching times obtained from the numerical integration of Eq.~(\ref{eq:tsw}) (lines) fit the data obtained from the solution of Eq.~(\ref{eq:sLLGS}) (symbols) very well for the two magnetic fields considered here. 

\subsection{Oscillation Dynamics}~\label{sec:oscillation}
\begin{figure}[ht!]
  \centering
  \includegraphics[width = 0.85\columnwidth, clip = true, trim = 0mm 0mm 0mm 0mm]{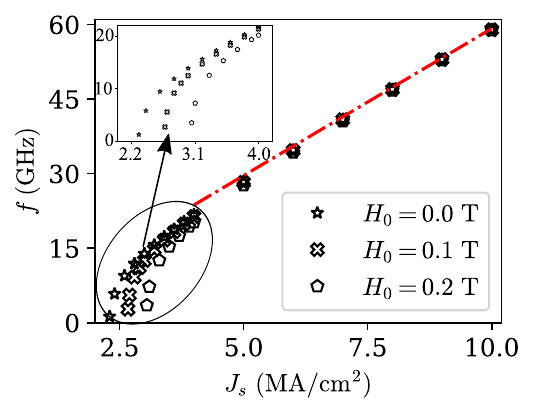}
  \caption{Oscillation frequency as a function of the applied spin current, $J_s$, and external magnetic field, $H_0$, for a thin film of Mn$_3$Sn under tensile strain. In each case, the applied current is above the respective $J_s^\mathrm{th2}$. 
  The dash-dotted red line represents $f = \frac{\gamma}{2\pi\alpha M_s}\frac{\hbar}{2e}\frac{J_s}{d_a}$ and fits the numerical data very well for large $J_s$. 
  The figure in the inset shows the variation in $f$ with $J_s$ for smaller input currents.} 
  \label{fig:frequency}
\end{figure}

For $H_0 > 0$ and $J_s^\mathrm{th2} \leq |J_s|$, \textcolor{black}{the SOT overcomes the maximum of the torques due to the internal and external magnetic fields, irrespective of the initial state or the direction of input current, resulting in $|\dot{\vb{m}}| > 0$ and $|\dot{\vb{m}}_i| > 0$.}
Consequently, \textcolor{black}{the magnetic octupole moves away from its initial stable state, crosses the barrier at $\varphi_\mathrm{oct} = 360^\circ$,} and oscillates between the two energy wells with frequencies ranging between 100's of MHz to {10s of} GHz, as shown in Fig.~\ref{fig:frequency}.
\textcolor{black}{Similar to the case of deterministic switching dynamics, $|m_z|$ increases till $\dot{\vb{m}}$ reaches zero due to the intrinsic damping, following which $|m_z|$ decreases. Here, however, $|m_z|$ increases again as $\dot{\vb{m}}$ increases, owing to the different direction of the torques in each energy well.
Therefore, $m_z$ exhibits two peaks of varying magnitude in each oscillation cycle, as shown clearly in the inset of Fig.~\ref{fig:phi_oct_osc1}(b). The higher peak occurs after the octupole moment crosses the energy barrier at $\varphi_\mathrm{oct} = 360^\circ$ whereas the lower peak occurs after it crosses the energy barrier at $\varphi_\mathrm{oct} = 180^\circ$.}

\begin{figure}[ht!]
  \centering
  \includegraphics[width = 0.85\columnwidth, clip = true, trim = 2mm 2mm 0mm 0mm]{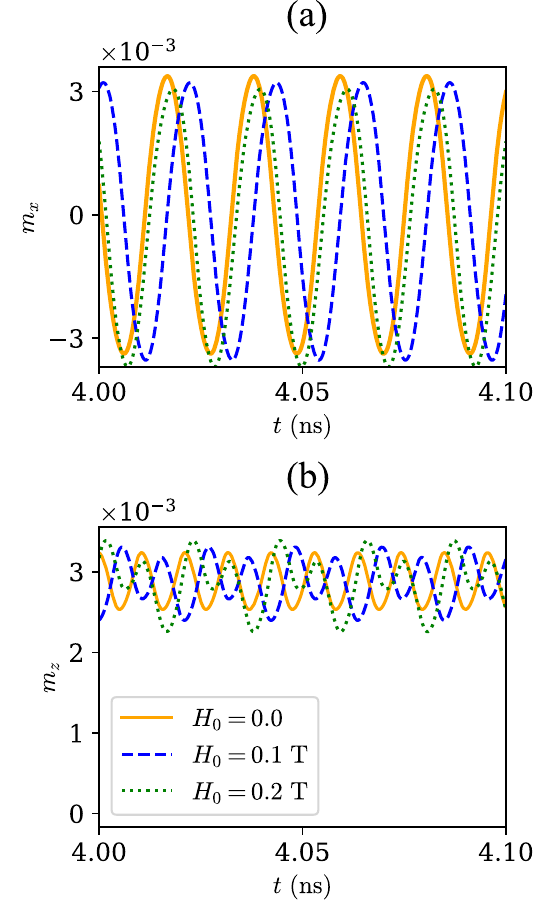}
  \caption{{(a) The x-component and (b) z-component of the average magnetization vector, $\vb{m}$, as functions of time, for large current $J_s = 8 ~\mathrm{MA/cm^2}$ and three different values of $H_0$. 
  Although the average oscillation frequency is evaluated to be the same for such large current, both $m_x$ and $m_z$ show the effect of $H_0$.}
  } 
  \label{fig:mx_mz_J8e10}
\end{figure}

{For medium to large currents,} the oscillation frequency, $f$, is almost independent of $H_0$ and increases linearly with $J_s$, as shown in Fig.~\ref{fig:frequency}.
The dash-dotted red line, which corresponds to $f = \frac{\gamma}{2\pi\alpha M_s}\frac{\hbar}{2e}\frac{J_s}{d_a}$, represents this behavior clearly.
For such $J_s$, $u_i$ increases to larger values. \textcolor{black}{Consequently, the effect of the out-of-(Kagome)-plane exchange interaction on the dynamics is expected to become significant while that of the in-plane anisotropy and $H_0$ is expected to reduce.}
The x- and z-components of $\vb{m}$ as functions of time, for $J_s = 8~\mathrm{MA/cm^2}$ and three different magnetic fields, are shown in Fig.~\ref{fig:mx_mz_J8e10}. Although the average frequency in the three cases is the same, there are subtle differences in the magnetization dynamics, owing to the symmetry-breaking magnetic field.
\textcolor{black}{In particular, $m_z$ is symmetric only for $H_0 = 0$ while it shows the expected asymmetry for non-zero $H_0$. Since the difference between the barrier heights increases with an increase in $H_0$ (Fig.~\ref{fig:energy}), the asymmetry in $m_z$ is more prominent for $H_0 = 0.2~\mathrm{T}$. A small asymmetry can also be observed in the sinusoidal $m_x$, where the magnitude in the negative (positive) x-direction increases (decreases) with $H_0$, owing to $\vb{H}_a$ being along the negative x-direction. 
Although not shown here, non-zero magnetic fields have negligible effect on the y-component of $\vb{m}$.}

\begin{figure}[ht!]
  \centering
  \includegraphics[width = 0.85\columnwidth, clip = true, trim = 2mm 2mm 0mm 0mm]{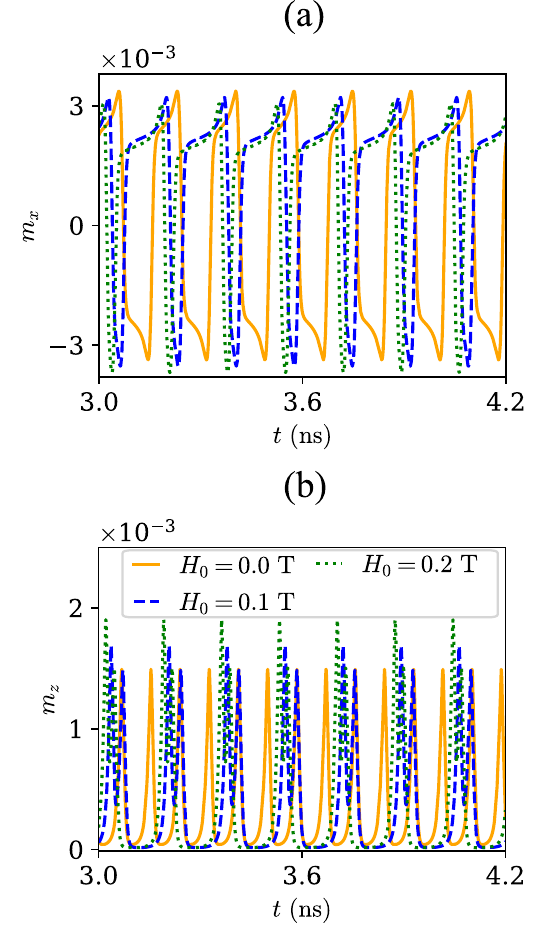}
  \caption{{(a) The x-component and (b) z-component of the average magnetization vector as functions of time at small current ($J_s \simeq J_s^{th2}$) for two different values of $H_0$. 
  The effect of the in-plane effective anisotropy is evident from the shape of $m_x$. 
  For non-zero $H_0$, the $m_z$ shows two spikes of varying amplitude due to energy barriers of different heights.} 
  } 
  \label{fig:mx_mz_JsJsth}
\end{figure}

For small currents, on the other hand, Fig.~\ref{fig:frequency} shows that $f$ increases non-linearly with $J_s$ \textcolor{black}{and depends on $H_0$.}
This dependence \textcolor{black}{of the oscillation dynamics on magnetic fields} can also be observed from Fig.~\ref{fig:mx_mz_JsJsth}, which shows the dynamics of $m_x$ and $m_z$ for three different magnetic fields near their respective threshold currents. 
\textcolor{black}{The strong in-plane anisotropy leads to non-sinusoidal $m_x$, unlike the case of large $J_s$ (Fig.~\ref{fig:mx_mz_J8e10}(a)). It also leads to a spike-like dynamics of $m_z$, where each oscillation of the magnetic octupole moment is accompanied by two spikes.
Non-zero magnetic field breaks the symmetry of the system leading to asymmetric profiles of $m_x$ and $m_z$.
The spikes in $m_z$ are equally spaced in time for $H_0 = 0~\mathrm{T}$. On the other hand, for $H_0 > 0$, the two spikes of each oscillation cycle are close to each other but far from those of the previous or next cycle. This is mainly due to the varying effects of the torque due to $\vb{H}_a$ as the magnetic octupole traverses the two energy wells.
}

Finally, in the non-linear regime, higher $H_0$ leads to lower $f$, at a fixed $J_s$, as depicted clearly in the inset of Fig.~\ref{fig:frequency}.
This is because the barrier height at $\varphi_\mathrm{oct} = 360^\circ$ increases with an increase in $H_0$, thereby requiring higher input energy in order to achieve the same oscillation frequency.
As current-driven oscillations are accompanied by large $m_z$, a strong exchange field along the z-direction affects the dynamics. 
However, since such an exchange energy interaction is not included in Eq.~(\ref{eq:energy_density_oct}), and consequently in Eq.~(\ref{eq:m_dot1}), our model cannot be used to obtain a unified model of $f$ as a function of $H_0$ and $J_s$. 

\vspace{-10pt}
\section{Effect of Damping}\label{sec:damping}
\vspace{-5pt}
The results presented in this work, so far, correspond to a Gilbert damping constant of $\alpha = 0.003$, which has previously been used for numerical analysis in Refs.~[\citen{tsai2020electrical, dasgupta2022tuning, shukla2023order}]. 
On the other hand, a lower damping constant of $\alpha = 0.0007$ was used in other recent works.~\citep{pal2022setting, higo2022perpendicular}
In particular in Ref.~[\citen{higo2022perpendicular}], it was shown through numerical simulations that for $\alpha = 0.0007$, as compared to $\alpha = 0.003$, the lower limit of external magnetic field for deterministic switching was a non-zero value. 
That is, for low values of $H_0 \gtrsim 0$, the \textcolor{black}{magnetic octupole moment} cannot be deterministically switched. Instead, it exhibits either a stationary steady-state or chiral oscillation depending on the magnitude of the input current. 
This behavior is distinct from that presented in Figs.~\ref{fig:final_state} and~\ref{fig:threshold_current}.
Therefore, we numerically investigate the dependence of the final steady-states and the threshold currents on $\alpha$. 

\begin{figure*}[ht!]
  \centering
  \includegraphics[width = 0.85\textwidth, clip = true, trim = 0mm 0mm 0mm 0mm]{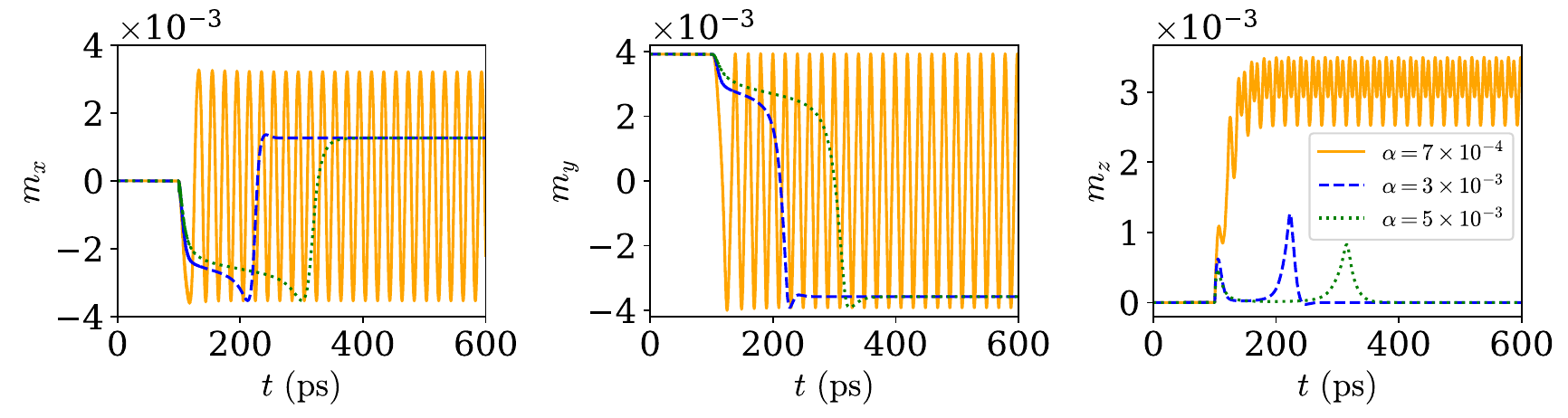}
  \caption{The three components of the average magnetization, $\vb{m}$, as a function of time for three different values of the Gilbert damping constant, $\alpha$.
  Here, $H_0 = 0.1~\mathrm{T}$ and $J_s = 2~\mathrm{MA/cm^2}$ are both applied to the equilibrium state of Fig.~\ref{fig:equilibrium}(c) at $t = 100~\mathrm{ps}$. 
  In the case of $\alpha = 0.003$ and $0.005$ the magnetic octupole moment switches to the same final steady-state in the other energy well. Switching time increases with $\alpha$. On the other hand, for $\alpha = 0.0007$, chiral oscillations with large $m_z$ are observed.} 
  \label{fig:avg_mag}
\end{figure*} 
\begin{figure*}[ht!]
  \centering
  \includegraphics[width = 0.85\textwidth, clip = true, trim = 0mm 70mm 0mm 69mm]{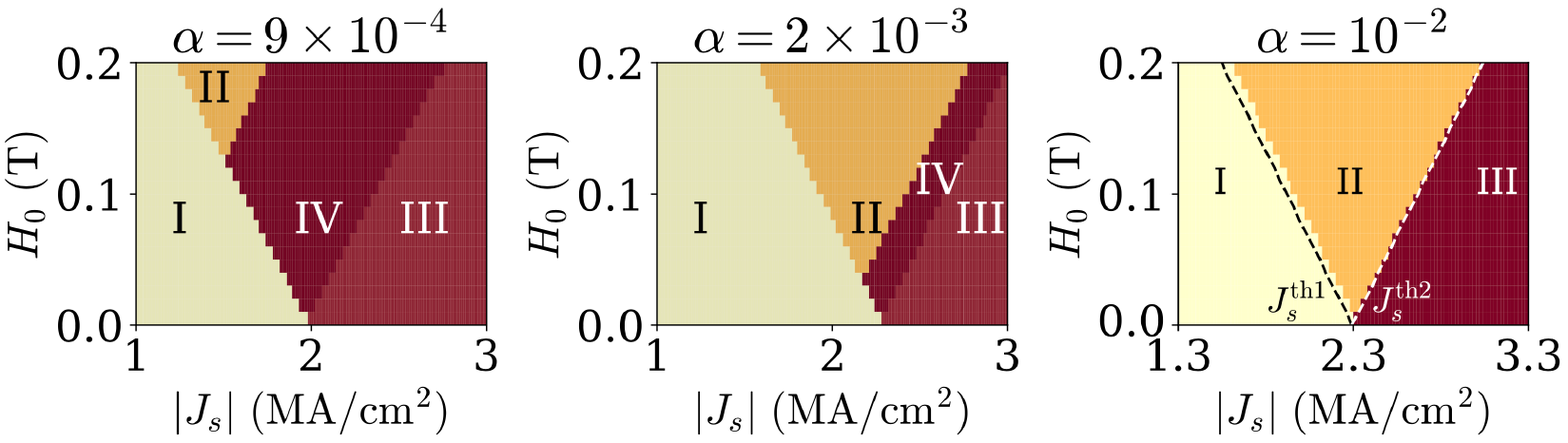}
 \caption{Final steady-state as a function of the input current $J_s$ and applied magnetic fields for different Gilbert damping constant.
  Here, I and II represent the case of no switching and switching while both III and IV
  correspond to \textcolor{black}{chiral oscillation}. The black dashed lines superimposed on the phase space, in the case of $\alpha = 0.01$, are the two threshold currents, as discussed in the case of $\alpha = 3 \times 10^{-3}$.} 
  \label{fig:final_state2}
\end{figure*} 
Figure~\ref{fig:avg_mag} shows the three components of the average magnetization vector, $\vb{m}$, for different values of $\alpha$ but the same values of $H_0$ and $J_s$. 
In the case of $\alpha = 5 \times 10^{-3}$, for $J_s = 2~\mathrm{MA/cm^2}$ and $H_0 = 0.1~\mathrm{T}$ applied at $t = 100~\mathrm{ps}$, the AFM magnetization switches to a steady state in the other energy well. 
This is signified by a change in the sign of $m_y$. As shown by the dashed blue curve and dotted green curve, the final steady-state for $\alpha = 5 \times 10^{-3}$ is exactly same as that obtained for $\alpha = 3 \times 10^{-3}$. The switching time, however, is longer in the case of higher damping since $t_\mathrm{sw}$ is directly proportional to $\alpha$ (Eq.~(\ref{eq:tsw})).
On the other hand, for the case of lower damping, namely $\alpha = 7 \times 10^{-4}$, the \textcolor{black}{magnetic octupole moment}
exhibits \textcolor{black}{chiral oscillations} when $J_s = 2~\mathrm{MA/cm^2}$ and $H_0 = 0.1~\mathrm{T}$ are applied at $t = 100~\mathrm{ps}$. 
This suggests that for lower damping, the threshold currents are lower than those predicted by Eq.~(\ref{eq:threshold}); and could be dependent on $\alpha$.
It can also be observed that for $\alpha = 7 \times 10^{-4}$, the oscillating $m_z$ is rather large.
This suggests that the out-of-(Kagome)-plane exchange interaction plays a major role in the oscillation dynamics, similar to that discussed in Section~\ref{sec:oscillation}.

\begin{figure}[ht!]
  \centering
  \includegraphics[width = 0.85\columnwidth, clip = true, trim = 0mm 0mm 0mm 0mm]{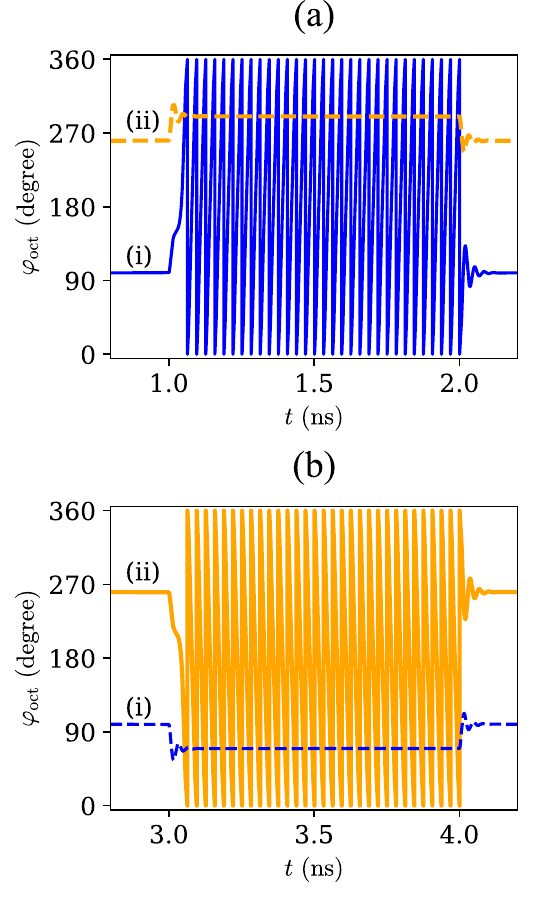}
  \caption{{Steady-state dynamics of the magnetic octupole moment as a function of time for $\alpha = 9 \times 10^{-4}$.
  (i) and (ii): Possible ground states for $H_0 = 0.1~\mathrm{T}$ in the negative x-direction.
  (a) The pulse is turned on at $t = 1~\mathrm{ns}$ to $J_s = 1.72~\mathrm{MA/cm^2}$ and turned off at $t = 2~\mathrm{ns}$. Steady-state oscillations for (i) but not for (ii).
  (b) The pulse is turned on at $t = 3~\mathrm{ns}$ to $J_s = -1.72~\mathrm{MA/cm^2}$ and turned off at $t = 4~\mathrm{ns}$. Steady-state oscillations for (ii) but not for (i).}  
  } 
  \label{fig:phi_oct_osc3}
\end{figure}

To further elucidate this dependence of the dynamics on $\alpha$, we present the phase space of the steady-states as a function of $J_s$ and $H_0$ for various $\alpha$ values in Fig.~\ref{fig:final_state2}.
Notably, the phase space analysis reveals an additional dynamical regime, labeled as IV, for $\alpha = 9 \times 10^{-4}$ and $2 \times 10^{-3}$, alongside the three previously identified regimes observed for $\alpha = 3 \times 10^{-3}$, and also found in the case of $\alpha = 0.01$.
As highlighted earlier for $\alpha = 0.003$, region I represents stationary steady-states in the proximity of the initial ground states, for $\varphi_\mathrm{oct}^\mathrm{init} = 90^\circ$ ($\varphi_\mathrm{oct}^\mathrm{init} = 270^\circ$) and $J_s > 0$ ($J_s < 0$). Region II, on the other hand, corresponds to deterministic switching between the two stable states, while the magnetic octupole moment exhibits chiral oscillations in region III, \textcolor{black}{regardless of the initial state or the direction of $J_s$, as long as $|J_s|$ exceeds the highest threshold current.}
Interestingly, in region IV, chiral oscillations occur for $\varphi_\mathrm{oct}^\mathrm{init} = 90^\circ$ ($\varphi_\mathrm{oct}^\mathrm{init} = 270^\circ$) and $J_s > 0$ ($J_s < 0$), while only stationary states are observed for $\varphi_\mathrm{oct}^\mathrm{init} = 270^\circ$ ($\varphi_\mathrm{oct}^\mathrm{init} = 90^\circ$). This scenario is vividly depicted in Fig.~\ref{fig:phi_oct_osc3} for $\alpha = 9 \times 10^{-4}$, $|J_s| = 1.72~\mathrm{MA/cm^2}$ and $H_0 = 0.1~\mathrm{T}$.

In contrast to the scenario with high damping, where the three regimes were distinguished by two threshold currents, the presence of four regimes in the low $\alpha$ case is marked by three distinct threshold currents.
Although all the three threshold currents seem to scale linearly with $H_0$ for $\alpha \lesssim 2\times 10^{-3}$, it can be clearly observed that they depend on the Gilbert damping constant, unlike Eq.~(\ref{eq:threshold}).
Moreover, for low values of $H_0$, deterministic switching is not possible; instead the \textcolor{black}{magnetic octupole moment} can only exhibit oscillation dynamics above the threshold current.
Deterministic switching between the two stable states of PMA Mn$_3$Sn becomes feasible again for larger values of $H_0$. 
This lower limit of $H_0$ for deterministic switching dynamics decreases as $\alpha$ increases. Although further analytic investigation is required to understand the dependence of the dynamics on the damping constant, we suspect that in region I the net input energy is low; therefore, the \textcolor{black}{magnetic octupole moment} cannot overcome the barrier at $180^\circ$. On the other hand, in region III the net input energy is very high such that \textcolor{black}{magnetic octupole moment} can exhibit sustained oscillations. In region IV, for low $H_0$, the barrier at $180^\circ$ is lowered which enables deterministic switching to the other energy well. However, the low damping of the system possibly does not dissipate \textcolor{black}{enough} energy of the \textcolor{black}{magnetic octupole moment}, and therefore, it goes over the barrier at $360^\circ$ due to its inertia. This leads to sustained oscillations. For higher fields, the barrier at $360^\circ$ becomes significantly large \textcolor{black}{and the magnetic octupole cannot overcome it}, such that deterministic switching becomes possible.  
Finally, we found that for $\alpha = 0.01$ the analysis presented in Section~\ref{sec:SOT} holds true. 
Since this analysis is true for both $\alpha = 3 \times 10^{-3}$ and $\alpha = 10^{-2}$, it is applicable for all other values of damping constants between them.

\vspace{-10pt}
\section{Conclusion}
\vspace{-5pt}
Mn$_3$Sn is a metallic antiferromagnet that shows large AHE, ANE, and MOKE signals. In addition, the octupole states can be detected via TMR in an all-antiferromagnetic tunnel junction comprising two layers of Mn$_3$Sn with an insulator layer sandwiched between them.
Bulk Mn$_3$Sn has a $120^\circ$ anti-chiral structure, however, a competition between the local anisotropy and the DMI leads to the existence of a small net magnetization which is six-fold degenerate. Application of strain to bulk Mn$_3$Sn reduces its symmetry from six-fold to two-fold degenerate, and provides a way to control the strength of the net magnetization as well as that of the AHE signal.
In this work, we analyzed the case of both uniaxial compressive and tensile strains, and discussed the dependence of the magnetic octupole \textcolor{black}{moment} on the strain as well as on the external field. 
Since recent experiment reported tensile strain in epitaxial Mn$_3$Sn grown on (110)[001] MgO substrate, we numerically and analytically explored the field-assisted SOT driven dynamics in monodomain Mn$_3$Sn with tensile strain.
We found that the \textcolor{black}{magnetic octupole moment} exhibits either a stationary state or chiral oscillations in the absence of a symmetry-breaking field.
On the other hand, when an external field is applied, in addition to the stationary state and chiral oscillations, the \textcolor{black}{magnetic octupole moment} can also be deterministically switched between the two stable states for a range of currents. 
We derived an effective equation which accurately predicts the stationary states in both the energy wells. We also derived simple analytic expressions of the threshold currents and found them to agree very well against the numerical results for small external magnetic fields.
We obtained functional form of the switching time as a function of the material parameters and the external stimuli and found it to match very well against numerical data.
The frequency of chiral oscillations, which can be tuned from 100s of MHz to {10s of GHz} range, was found to vary non-linearly closer to the threshold current and linearly for larger input currents.
Further, through numerical simulations, we showed that the order dynamics is dependent on the Gilbert damping for lower values of $\alpha$. 
For the sake of a complete picture, we also explored the field-assisted switching dynamics in thin films of Mn$_3$Sn with \textcolor{black}{no strain as well as} compressive strain, and presented the relevant results in the supplementary document.
We expect the insights of our theoretical investigation to be useful to both theorists and experimentalists in their exploration of the interplay of field-assisted SOT and the order dynamics in Mn$_3$Sn, and further benchmarking the device performance.

\section*{Supplementary Material}
See supplementary material for the perturbative analysis of the ground state, the SOT-driven dynamics of thin films of Mn$_3$Sn with compressive strain as well as no strain, and a brief discussion of the AHE and TMR detection schemes. 

\section*{Data Availability Statement}
The data that support the findings of this study are available from the corresponding author upon reasonable request.

\begin{acknowledgments}
This research was supported by the NSF through the University of Illinois at Urbana-Champaign Materials Research Science and Engineering Center DMR-1720633.
\end{acknowledgments}


\nocite{}
\bibliography{aipsamp}

\end{document}